\documentclass[a4paper,11pt]{article}

\pdfoutput=1
\usepackage{amsmath,amssymb,cite,tabularx,graphicx,color,cancel,xspace,flafter,ifpdf,todonotes,subcaption,microtype,hyperref}

\newcommand{\HEJ}{{\tt HEJ}\xspace}
\newcommand{\HIGHEJ}{\emph{High Energy Jets}\xspace}

\newcommand{\as}{\ensuremath{\alpha_s}\xspace}

\newcommand{\madgraph}{\texttt{MadGraph5\_aMC@NLO}\xspace}
\def\spa#1.#2{\left\langle#1\,#2\right\rangle}
\def\spb#1.#2{\left[#1\,#2\right]}
\def\spaa#1.#2.#3{\langle\mskip-1mu{#1}
                  | #2 | {#3}\mskip-1mu\rangle}
\def\spbb#1.#2.#3{[\mskip-1mu{#1}
                  | #2 | {#3}\mskip-1mu]}
\def\spab#1.#2.#3{\langle\mskip-1mu{#1}
                  | #2 | {#3}\mskip-1mu\rangle}
\def\spba#1.#2.#3{\langle\mskip-1mu{#1}^+
                  | #2 | {#3}^+\mskip-1mu\rangle}
\def\spav#1.#2.#3{\|\mskip-1mu{#1}
                  | #2 | {#3}\mskip-1mu\|^2}
\def\jc#1.#2.#3{j^{#1}_{#2#3}}

\title{\begin{normalsize}
\begin{flushright}
DCPT/18/66, IPPP/18/33, MCnet-18-10
\end{flushright}
\end{normalsize}
\vspace*{1cm}Higgs-boson plus Dijets:\\Higher-Order Matching for High-Energy Predictions}
\author{Jeppe~R.~Andersen$^{a}$, Tuomas Hapola$^{a}$, Marian Heil$^{a}$, Andreas Maier$^{a}$\\ and
  Jennifer~M.~Smillie$^{b}$\\\mbox{}\\
  $^a$ Institute for Particle Physics Phenomenology,\\University of Durham,
  South Road, Durham DH1 3LE, UK\\
  $^b$ Higgs Centre for Theoretical Physics, University of Edinburgh,\\
  Peter Guthrie Tait Road, Edinburgh EH9 3FD, UK.}

\begin{document}
\maketitle
\begin{abstract}
  Several important processes and analyses at the LHC are sensitive to
  higher-order perturbative corrections beyond what can currently be
  calculated at fixed order. The formalism of \HIGHEJ (\HEJ) calculates the
  corrections systematically enhanced for a large ratio of the centre-of-mass
  energy to the transverse momentum of the observed jets. These effects are
  relevant in the analysis of e.g.~Higgs-boson production in association with
  dijets within the cuts devised to enhance the contribution from Vector
  Boson Fusion (VBF).

  \HEJ obtains an all-order approximation, based on logarithmic corrections
  which are
  matched to fixed-order results in the cases where these can be readily
  evaluated. In this paper we present an improved framework for the matching
  utilised in \HEJ, which for merging of tree-level results is mathematically
  equivalent to the one used so far. However, by starting from events
  generated at fixed order and supplementing these with the all-order summation,
  it is computationally simpler to obtain matching to calculations of high
  multiplicity.

  We demonstrate that the impact of the higher-multiplicity matching on
  predictions is small for the gluon-fusion (GF) contribution of Higgs-boson
  production in association with dijets in the VBF-region, so perturbative
  stability against high-multiplicity matching has been achieved within
  \HEJ. We match the improved \HEJ prediction to the inclusive
  next-to-leading order (NLO) cross section and compare to pure NLO in
  the $h\to\gamma\gamma$ channel with standard VBF cuts.
\end{abstract}

\newpage
\boldmath
\tableofcontents
\unboldmath
\section{Introduction}
\label{sec:Introduction}
Fixed-order perturbation theory delivers a good description of inclusive
rates of collider-processes involving jets; however, logarithmic corrections of
various origins are important for observables in different regions of phase space. 
For example, the detailed description of
the dependence of the cross-section on jet-sizes $R$ receives systematic
logarithmic perturbative corrections of the type
$\alpha_s^n \ln^n 1/R$\cite{Dasgupta:2014yra,Dasgupta:2016bnd}. These
logarithms are controlled by DGLAP-like evolution equations, which also
govern the formalism of parton
showers\cite{Bellm:2015jjp,Sjostrand:2014zea,Gleisberg:2008ta}. 
While corners of phase space
characterised by large ratios of transverse scales are well described by the
parton-shower formalism, measurements at D0 at 1.96~TeV\cite{Abazov:2013gpa}
and ATLAS at 7~TeV\cite{Aad:2014qxa} indicate clearly that even when matched
with fixed-order matrix elements, the parton showers do not describe well the
regions of large invariant mass or large rapidity spans of the jet
systems. This region of phase space is of particular interest in the process
$pp\to Hjj$, with contributions (at Born level) of $\alpha_w^2$ through weak
boson fusion and $\alpha_s^4$ through gluon-fusion (GF). It is reasonable to
distinguish the two contributions to the same final state, since the quantum
interference is negligible\cite{Andersen:2007mp,Bredenstein:2008tm,Dixon:2009uk}. The impact of the radiative
corrections to each process is rather different though; in particular, the
$t$-channel colour octet exchange in the gluon-fusion process leads to
increased jet-activity\cite{Dokshitzer:1991he}, which allows for a
distinction of the production mechanism within the phase space populated by
weak boson fusion. The two jets in weak boson fusion are often separated by a
large invariant mass and rapidity span. This is the phase-space region where
the perturbative corrections for the QCD processes contain logarithms of
$\log(\hat s_{jj}/p_t^2)$ from Balitsky-Fadin-Kuraev-Lipatov
(BFKL)\cite{Fadin:1975cb,Kuraev:1976ge,Kuraev:1977fs,Balitsky:1978ic}. These
logarithms are contained within the formalism of \HIGHEJ, where the
systematic treatment is obtained by a power-expansion of the scattering
matrix element in $\hat s/p_t^2$\cite{Andersen:2008ue,Andersen:2008gc}. The
first sub-leading corrections were presented for the process of $Hjj$ in
Ref.\cite{Andersen:2017kfc} by calculating the leading behaviour of certain
sub-leading processes. This constitutes control of a well-defined set of NLL
BFKL logarithms within \HEJ. These logarithms drive the pattern
of further emissions from the QCD process\cite{Dokshitzer:1991he}, which will
allow for a better discrimination between the GF and VBF processes than what
could be performed by investigating the dynamics of just two jets
in the event.

The formalism of \HEJ captures leading logarithmic terms to processes with at
least two jets at large partonic centre-of-mass energy, of the form
$\as^k \ln(\hat s/p_\perp^2)^k\simeq \as^k \Delta y_{j_f j_b}^k$, where
$\Delta y_{j_f j_b}$ is the rapidity-difference between the jets forward and
backward in rapidity. The systematic treatment of these terms is based on a
logarithmic all-order expansion point-by-point in phase space of the leading
virtual corrections to all orders, combined with a power-expansion (in
$\hat s/p_t^2$) of the square of the tree-level matrix elements, again
point-by-point in the $n$-particle phase space. Upon integration, the leading
power-expansion of the square of the matrix elements ensure the appropriate
logarithmic accuracy of cross sections. The contributions for $n>2$ are
numerically integrated over phase space, allowing for detailed jet clustering
and event analyses.

Within the formalism of \HEJ, the $m$-jet rates entering each prediction are
matched to tree-level accuracy point-by-point in phase space, by the
following procedure for mapping the $n$-parton resummation phase space point
into a $m$-parton tree-level phase space point, described in more detail in
Ref.\cite{Andersen:2011hs} and Section~\ref{sec:matching}:-
\begin{enumerate}
\item cluster the $n$-parton phase space point into jets with a chosen jet
  algorithm and jet $p_t$-threshold (e.g.~anti-kt clustering, with a
  threshold of 30~GeV)
\item remove the partons not forming part of the hard jets from the event,
  and distribute the sum of their transverse momenta onto the hard jets
\item adjust the energy and longitudinal
  momentum of each jet such that it is on-shell, while keeping their rapidities fixed
\item adjust the momenta of the incoming partons such that energy and
  momentum conservation is restored
\end{enumerate}
The result of this procedure is a set of momenta for which the on-shell
$m$-jet tree-level matrix element can be evaluated. This in turns allows for
the weight of the generated event to be reweighted to full tree-level $m$-jet
accuracy, thus obtaining full tree-level accuracy up to the multiplicity for
which the tree-level matrix elements can be evaluated in reasonable
time. This method for matching the all-order results to fixed-order
high-multiplicity matrix elements has been used for matching all results
obtained with \HEJ: jets, and $\gamma$, $Z$, $W$ plus at least two jets
with matching up to 4 jets, and $H$ with at least two jets, with matching up
to 3 jets. 

The matching procedure described here can thus be viewed as \emph{merging}
the results of fixed-order calculations by use of the power-expanded matrix
elements of \HEJ coupled with the logarithmic virtual corrections, similarly
to the CKKW-L--method\cite{Catani:2001cc,Lonnblad:1992tz} of using the
logarithmic accuracy of a shower-algorithm to merge fixed-order cross
sections of varying multiplicity. This paper will present a complete
reformulation of the procedure for merging and all-order summation. With the
same input (as in use of the same matrix elements to the same order), the
results are unchanged, but the new procedure for obtaining the all-order
results and the merging will allow for merging results beyond tree-level, and
will be computationally much more efficient. 

In Section~\ref{sec:matching} we describe the original mechanism for matching
leading-order samples within \HEJ before a detailed discussion of the new
formulation.  This includes both analytical aspects and practical aspects of
implementation.  In Section~\ref{sec:results} we study the results obtained in
the new formalism in the context of Higgs boson plus dijets in three
studies. Firstly we confirm that when matching to fixed order samples is limited
to a maximum of three jets, we find consistent results with the previous
formalism.  Secondly, we study the impact of increasing the multiplicity in
the fixed-order samples, now possible for the first time.  Thirdly, we compare
the matched all-order results of \HEJ with those obtained at next-to-leading
order accuracy.  In Section~\ref{sec:summary}, we conclude with a final discussion.



\section{Matching}
\label{sec:matching}


In the original formulation, the cross sections within \HEJ are calculated by explicitly
constructing the all-order result by first generating a $2\to n+l$ kinematic
point for each number of partons $n=2,\ldots,N$, where $N$ is chosen
sufficiently large (in practice around 22), and $l$ describes the non-partonic particles produced,
e.g.~$Z,W,H$ or their decay products. In order to simplify the notation
we will only discuss the purely partonic case. Likewise, we will
restrict our discussion to the leading-logarithmic
contribution. Note that all our arguments apply equally to the more
general scenario. We demonstrate this by showing results for
the production of a Higgs boson in association with at least two jets,
including recently computed sub-leading
corrections~\cite{Andersen:2017kfc}.

The high-energy limit is dominated by Fadin-Kuraev-Lipatov (FKL)
configurations, where two partons scatter in such a way that there is no
radiation outside the rapidity range spanned by the scattering partons
and only gluons are emitted inside this range. To ensure the high-energy
limit applied is valid, it is required that the extremal (in rapidity)
partons are perturbative (hard in terms of transverse momentum), and are
members of the extremal jets. The transverse momenta of the remaining
partons are all generated down to effectively $0$~GeV, technically to a
very small scale of order 200~MeV (which can be varied), below which
there is perfect cancellation between the subtraction terms (used in the
organisation of the cancellation of the IR
divergences~\cite{Andersen:2017kfc}) and the real-emission terms. The
matching to LO accuracy for all $m$-jet rates, $m\le n$, is then
obtained by first projecting the kinematics of the generated all-order
events into Born kinematics according to the number of hard jets as described in
the previous section. The event weight is multiplied with a ratio of the square of
the full Born-level matrix element to the \HEJ approximation of the
same. The cross section (and kinematic distributions) is then obtained
through the formula:-
\begin{align}
  \begin{split}
    \label{eq:resumdijetFKLmatched}
    \sigma_{2j}^\mathrm{resum, match}=&\sum_{f_1, f_2}\ \sum_{n=2}^\infty\
\int_{p_{1\perp}=p_{\perp,\mathrm{min}} }^{p_{1\perp}=\infty}
      \frac{\mathrm{d}^2\mathbf{p}_{1\perp}}{(2\pi)^3}\
\int_{p_{n\perp}=p_{\perp,\mathrm{min}}}^{p_{n\perp}=\infty}
      \frac{\mathrm{d}^2\mathbf{p}_{n\perp}}{(2\pi)^3}\
    \prod_{i=2}^{n-1}\int_{p_{i\perp}=\lambda}^{p_{i\perp}=\infty}
      \frac{\mathrm{d}^2\mathbf{p}_{i\perp}}{(2\pi)^3}\\
    &\times \ \mathbf{T}_y \prod_{i=1}^n 
      \left(\int \frac{\mathrm{d} y_i}{2}\right)\ \times \frac{\overline{|\mathcal{M}_{\mathrm{HEJ}}^{\text{reg}}(\{ p_i\})|}^2}{\hat s^2} 
    \times\ \sum_m \mathcal{O}_{mj}^e(\{p_i\})\ w_{m\mathrm{-jet}}^{\mathrm{LO}}\\
    &\times\ \ x_a f_{a,f_1}(x_a, Q_a)\ x_b f_{b,f_2}(x_b, Q_b)\ (2\pi)^4\ \delta^2\!\!\left(\sum_{i=1}^n
      \mathbf{p}_{i\perp}\right )\ \mathcal{O}_{2j}(\{p_i\}),
  \end{split}
\end{align}
where $|\mathcal{M}_{\mathrm{HEJ}}^{\text{reg}}(\{ p_i\})|^2$ is the square
of the regularised all-order matrix element within \HEJ for the $2\to n$
phase space point (see
Ref.\cite{Andersen:2017kfc} for further details), and
\begin{align}
  \label{eq:matchfact}
  w^{\mathrm{LO}}_{m\mathrm{-jet}}\equiv\frac{\overline{\left|\mathcal{M}_\text{LO}^{f_1f_2\to f_1g\cdots
          gf_2}\left(\left\{p^B_{\mathcal{J}_l}(\{p_i\})\right\}\right)\right|}^2}{\overline{\left|\mathcal{M}_{\text{LO,
  HEJ}}^{f_1f_2\to
          f_1g\cdots
          gf_2}\left(\left\{p^B_{\mathcal{J}_l}(\{p_i\})\right\}\right)\right|}^2}
\end{align}
is the ratio between the square of the matrix element evaluated at full
tree-level accuracy and within \HEJ for the state projected to tree-level
$2\to m$ kinematics described by the jet momenta
$\left\{p^B_{\mathcal{J}_l}(\{p_i\})\right\}$.
$\mathcal{O}_{mj}^e(\{p_i\})$ is the exclusive $m$-jet measure applied to the
generated event kinematics. $\mathbf{T}_y$ indicates rapidity ordering. The
limits of the integral over the transverse momentum of the extremal partons
combined with the two-jet measure is set to guarantee that the extremal
partons carry the dominant momentum of the extremal jets. We choose a
cut-off $p_{\perp,\mathrm{min}}$ corresponding to 90\% of the
transverse momentum of the respective extremal jet.\footnote{This is a
  slightly more sophisticated cut than that investigated in
  Ref.\cite{Andersen:2011hs,Andersen:2012gk,Andersen:2016vkp,Andersen:2017kfc},
  and ensures that the soft divergence which would be regulated at
  next-to-leading logarithmic accuracy of the extremal currents does not
  impact the result obtained with the leading-logarithmic currents even for
  jets at large transverse momentum.} We use here the phrase `kinematics of
the generated all-order event' to mean the $n$-parton kinematic point of the
resummation event sampled in Eq.~\eqref{eq:resumdijetFKLmatched}. In order to
match each $m$-jet rate to tree-level accuracy, each generated event in the
all-order phase-space is mapped to a $m$-jet tree-level kinematic point, and
requires an evaluation of the full $m$-jet matrix element.

The scale-variation of the normalisation of the cross sections is determined
by the tree-level matrix elements, and mostly unchanged by the leading
logarithmic (LL) high-energy resummation implemented in \HEJ. This could be
reduced by extending the reweighting factor $w_{m-\mathrm{jet}}$ to
next-to-leading order accuracy.  However, in order to do this, 
one would
have to integrate over all $m+1$ parton real emission phase space resulting
in a specific $m$-jet Born level kinematics. This would be prohibitively
time-consuming.  An opposite approach is to begin with fixed order samples of
exclusive jet rates and then merge these using \HEJ to generate all-order
results.  We demonstrate how to do this in the next subsection and find
significant benefits already at tree-level accuracy.  In particular, each phase
space point used for the tree-level matrix element maps into all the relevant
resummation phase space points which leads to fewer evaluations of the
tree-level matrix elements.  This in turn allows for matching to higher
multiplicity for a given CPU envelope.

\subsection{Supplementing Fixed Order Samples with HEJ Resummation}
\label{sec:suppl-fixed-order}


The reformulation of the resummation and matching should reproduce the
results of Eq.~\eqref{eq:resumdijetFKLmatched}. Starting from this equation,
we introduce a $\delta$-functional and an integration over the Born level
kinematics of the on-shell, reshuffled jets $\{j_B^i\}$ reconstructed from
the resummed kinematics. Eq.~\eqref{eq:resumdijetFKLmatched} is rewritten to
\begin{align}
    \label{eq:resumdijetFKLmatched2}
      \sigma&_{2j}^\mathrm{resum, match}=\sum_{f_1, f_2}\ \sum_m
      \prod_{j=1}^m\left(
        \int_{p_{j\perp}^B=0}^{p_{j\perp}^B=\infty}
        \frac{\mathrm{d}^2\mathbf{p}_{j\perp}^B}{(2\pi)^3}\ \int
        \frac{\mathrm{d} y_j^B}{2} \right) \
      (2\pi)^4\ \delta^{(2)}\!\!\left(\sum_{k=1}^{m}
        \mathbf{p}_{k\perp}^B\right)\nonumber\\
      &\times\ x_a^B\ f_{a, f_1}(x_a^B, Q_a^B)\ x_b^B\ f_{b, f_2}(x_b^B, Q_b^B)\
      \frac{\overline{\left|\mathcal{M}_\text{LO}^{f_1f_2\to f_1g\cdots
          gf_2}\big(\big\{p^B_j\big\}\big)\right|}^2}{(\hat {s}^B)^2}\nonumber\\
& \times \frac{w_{m\mathrm{-jet}}}{\overline{\left|\mathcal{M}_\text{LO}^{f_1f_2\to f_1g\cdots
          gf_2}\big(\big\{p^B_j\big\}\big)\right|}^2}\ \times
(2\pi)^{-4+3m}\ 2^m \nonumber\\
&\times\ \sum_{n=2}^\infty\
\int_{p_{1\perp}=p_{\perp,\mathrm{min}} }^{p_{1\perp}=\infty}
      \frac{\mathrm{d}^2\mathbf{p}_{1\perp}}{(2\pi)^3}\
 \int_{p_{n\perp}=p_{\perp,\mathrm{min}}}^{p_{n\perp}=\infty}
       \frac{\mathrm{d}^2\mathbf{p}_{n\perp}}{(2\pi)^3}\
     \prod_{i=2}^{n-1}\int_{p_{i\perp}=\lambda}^{p_{i\perp}=\infty}
       \frac{\mathrm{d}^2\mathbf{p}_{i\perp}}{(2\pi)^3}\ (2\pi)^4\ \delta^{(2)}\!\!\left(\sum_{k=1}^n
        \mathbf{p}_{k\perp}\right )\nonumber\\
     &\times \ \mathbf{T}_y \prod_{i=1}^n 
       \left(\int \frac{\mathrm{d} y_i}{2}\right)\ 
      \mathcal{O}_{mj}^e\
      \left(\prod_{l=1}^{m-1}\delta^{(2)}(\mathbf{p}_{\mathcal{J}_{l}\perp}^B -
       \mathbf{j}_{l\perp})\right)\  \left(\prod_{l=1}^m\delta(y^B_{\mathcal{J}_l}-y_{\mathcal{J}_l})\right)\\
     &\times x_a f_{a,f_1}(x_a, Q_a)\ x_b f_{b,f_2}(x_b, Q_b)\ \frac{\overline{|\mathcal{M}_{\mathrm{HEJ}}^{f_1 f_2\to f_1 g\cdots
            gf_2}(\{ p_i\})|}^2}{\hat s^2} \ \mathcal{O}_{2j}(\{p_i\})\nonumber\\
      &\times\ \frac{(\hat {s}^B)^2}{x_a^B\ f_{a,f_1}(x_a^B, Q_a^B)\ x_b^B\ f_{b,f_2}(x_b^B, Q_b^B)}.\nonumber
\end{align}
The first two lines are now the phase space integration over the LO matrix
element, which can be represented in terms of (potentially weighted) tree-level
events. Obviously, the Born-level partonic momenta are identical with
the Born-level jet momenta, i.e. $p^B_i \equiv p^B_{\mathcal{J}_i}$, so that

\begin{equation}
\label{eq:reweight_factor}  
\frac{w_{m\mathrm{-jet}}}{\overline{\left|\mathcal{M}_\text{LO}^{f_1f_2\to f_1g\cdots
          gf_2}\big(\big\{p^B_j\big\}\big)\right|}^2} =
    \overline{\left|\mathcal{M}_\text{LO, HEJ}^{f_1f_2\to f_1g\cdots
          gf_2}\big(\big\{p^B_j\big\}\big)\right|}^{-2}
\end{equation}
only depends on the Born-level \HEJ approximation to the matrix
element. Lines 4--6 are the integration of the \HEJ matrix elements over
all of the resummation phase space, which map onto the given
fixed-order kinematics. Finally, line 7 removes the factors introduced in
the first line of Eq.~\eqref{eq:resumdijetFKLmatched2} compared to
Eq.~\eqref{eq:resumdijetFKLmatched} in order to write the matching in
terms of a standard phase space integration over fixed-order PDFs and
matrix elements.

The $\delta$-functionals of the fifth line in
Eq.~\eqref{eq:resumdijetFKLmatched2} connect the reconstructed Born-level
kinematics with the kinematics of the jets arising from the resummation. The
algorithm devised for projecting the jet momenta of the resummation onto
Born-level kinematic gives\cite{Andersen:2011hs}
\begin{align}
  \label{eq:ptreassign}
  \mathbf{p}^B_{\mathcal{J}_{l\perp}} = \mathbf{j}_{l\perp} \equiv \mathbf{p}_{\mathcal{J}_{l}\perp} + \mathbf{q}_\perp \,
  \frac{|\mathbf{p}_{\mathcal{J}_{l}\perp}|}{P_\perp},
\end{align}
plus the constraint that the rapidities of the jets are kept fixed. Here
$\mathbf{p}^B_{\mathcal{J}_{l}}$ is the momentum of the fixed-order, matching
level jet, $\mathbf{q}_\perp$ is the sum of the transverse momenta of partons
outside jets after
resummation, which equals minus the transverse momentum of the jets after
resummation. $P_\perp$ is the scalar sum of the jet transverse
momenta after resummation. This algorithm can be straightforwardly
applied when the resummation event has been constructed, and had a
jet-clustering applied. If, however, we want to start from fixed-order
generated events, the algorithm needs to be inverted, such that all
resummation-momenta on the right-hand side of Eq.~\eqref{eq:ptreassign}
are explored for a given Born-level kinematic point.

\begin{figure}[tbp]
  \centering
   \includegraphics[width=.75\textwidth]{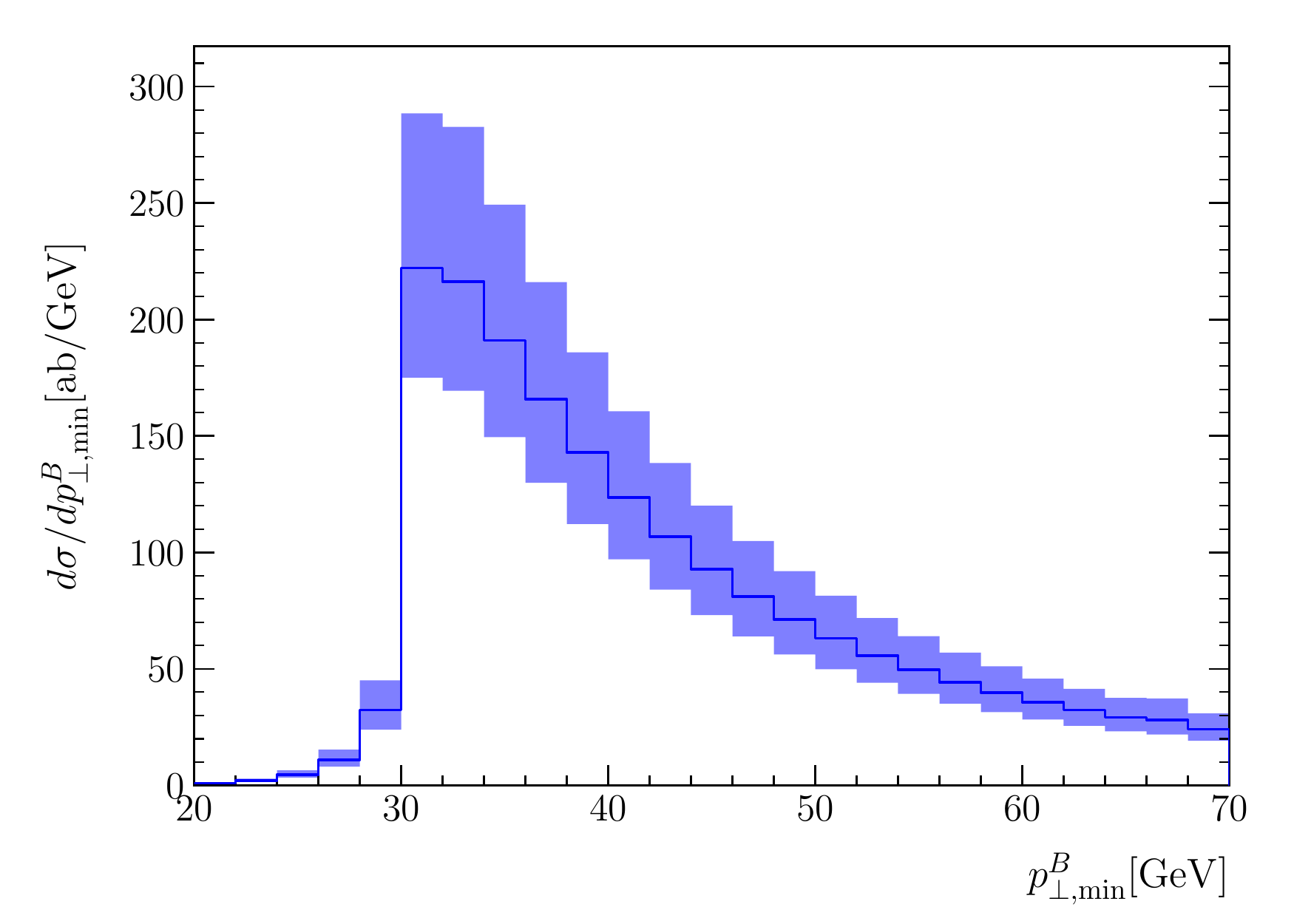}
   \caption{The distribution of the minimum transverse momentum of jets used
     in the matching, $p^B_{\perp, \mathrm{min}}$, for Higgs boson plus dijet
     production of transverse momenta larger than 30~GeV. The fact that the
     distribution falls off quickly below the jet analysis scale ensures the
     resummation phase space with a given minimum jet transverse momentum is
     covered by a fixed-order generation with a slightly smaller requirement
     on the jet transverse momentum. For example, a generation of
     fixed-order events with a minimum jet transverse momentum of 20~GeV is
     sufficient for an analysis requiring a transverse momentum of at least 30~GeV.}
   \label{fig:pjperpmin}
\end{figure}
While Eq.~\eqref{eq:resumdijetFKLmatched2} is mathematically equivalent to
Eq.~\eqref{eq:resumdijetFKLmatched}, it does not prove that the approach is
viable. The first challenge is to ensure that in fact, the integration over
the matching, or fixed-order phase space, in the first line of
Eq.~\eqref{eq:resumdijetFKLmatched2} does not actually extend to zero
transverse momentum of the matching jets. This would lead to a
divergence in the fixed-order cross section and invalidate the starting
point. In
Fig.~\ref{fig:pjperpmin} we investigate the
minimum transverse momentum of jets used in the matching for the evaluation
of fixed-order matrix elements. The plot shows
$d\sigma/d p^B_{\perp, \mathrm{min}}$, where $p^B_{\perp, \mathrm{min}}$
is the minimum jet transverse momentum used
in the merging with matrix elements (i.e.~the minimum transverse momentum in
the resulting on-shell Born level kinematics after reshuffling) for Higgs-boson
production in association with at least two jets with transverse momentum of
at least 30~GeV.
One sees that the matrix element sample needs to include events with a minimum
jet transverse momentum \emph{below} the final analysis scale --- but 
not \emph{too} far below. It is observed that this distribution gets broader, and that the
weight for small $p^B_{\perp, \mathrm{min}}$ is relatively more important, both for
larger rapidity spans, and if more hard jets are required (obviously these
two requirements are linked).

The next challenge is to generate all resummation kinematics corresponding to a
specific fixed-order or matching kinematics. This is not an obvious
switch to make: substituting a requirement on (N)LO kinematics to result in
a given Born level jet configuration with that of the full resummation event
resulting in a given Born level jet kinematics. The formalism will though be
much more computationally efficient, since the fixed-order matrix element is evaluated only
once for each fixed-order kinematic point. And furthermore, statistical
convergence can be ensured first at the fixed-order stage, before resummation
is attempted. 

\subsection{Phase Space Generation}
\label{sec:psp_gen}

In order to perform the resummation, we are tasked with the numerical
evaluation of the last four lines of
Eq.~\eqref{eq:resumdijetFKLmatched2}. In principle, we have to integrate
over the phase space of arbitrarily many further real emissions. This is
made feasible by the fact that for a given fixed-order configuration
with finite rapidity span, only a limited number of additional gluons actually lead to a
non-negligible contribution in the resummation. Still, the typical multiplicities in the
interesting region of large rapidity separations will be quite high and
we are required to inspect the corresponding high-dimensional phase
space carefully for an efficient integration. In the following, we discuss
how to construct an efficient importance sampling.

\subsubsection{Gluon Multiplicity}
\label{sec:psp_ng}

The typical number of extra emissions depends strongly on the rapidity
span of the underlying fixed-order event.
Let us, for example, consider a fixed-order FKL-type multi-jet
configuration with rapidities $y_{j_f},\,y_{j_b}$ of the most forward and
backward jets, respectively. By construction of the matching algorithm of Ref.\cite{Andersen:2011hs}, the jet multiplicity and
the rapidity of each jet are conserved when adding resummation. This
implies that additional hard radiation is restricted to rapidities $y$
within a region $y_{j_b} \lesssim y \lesssim y_{j_f}$. Within \HEJ, we require
the most forward and most backward emissions to be hard in order to
avoid divergences~\cite{Andersen:2011hs}, so this
constraint in fact applies to \emph{all} additional radiation.

To simplify the remaining discussion, let us remove the FKL rapidity
ordering
\begin{equation}
  \label{eq:remove_y_order}
  \mathbf{T}_y \prod_{i=1}^n\int \frac{\mathrm{d}y_i}{2} =
  \frac{1}{n!}\prod_{i=1}^n\int
  \frac{\mathrm{d}y_i}{2}\,,
\end{equation}
where all rapidity integrals now cover a region which is approximately
bounded by $y_{j_b}$ and $y_{j_f}$. Each of the $m$ jets has to contain at least
one parton; selecting random emissions we can rewrite the phase space
integrals as
\begin{equation}
  \label{eq:select_jets}
  \frac{1}{n!}\prod_{i=1}^n\int [\mathrm{d}p_i] =
  \left(\prod_{i=1}^{m}\int [\mathrm{d}p_i]\ {\cal J}_i(p_i)\right)
  \frac{1}{n_g!}\prod_{i=m+1}^{m+n_g}\int [\mathrm{d}p_i]
\end{equation}
with jet selection functions
\begin{equation}
  \label{eq:def_jet_selection}
  {\cal J}_i(p) =
  \begin{cases}
    1 &p\text{ clustered into jet }i\\
    0 & \text{otherwise}
  \end{cases}
\end{equation}
and $n_g \equiv n - m$. Here and in the following we use the short-hand
notation $[\mathrm{d}p_i]$ to denote the phase-space measure for parton
$i$. As is evident from Eq.~\eqref{eq:select_jets}, adding an extra emission
$n_g+1$ introduces a suppression factor $\tfrac{1}{n_g+1}$. However, the
additional phase space integral also results in an enhancement proportional
to $\Delta y_{j_f j_b} = y_{j_f} - y_{j_b}$. This is a result of the
rapidity-independence of the MRK limit of the integrand, consisting of the
matrix elements divided by the flux factor. Indeed, we observe that the
typical number of gluon emissions is to a good approximation proportional to
the rapidity separation and the phase space integral is dominated by events
with $n_g \approx \Delta y_{j_f j_b}$.

\begin{figure}
   \centering
   \includegraphics[width=.75\textwidth]{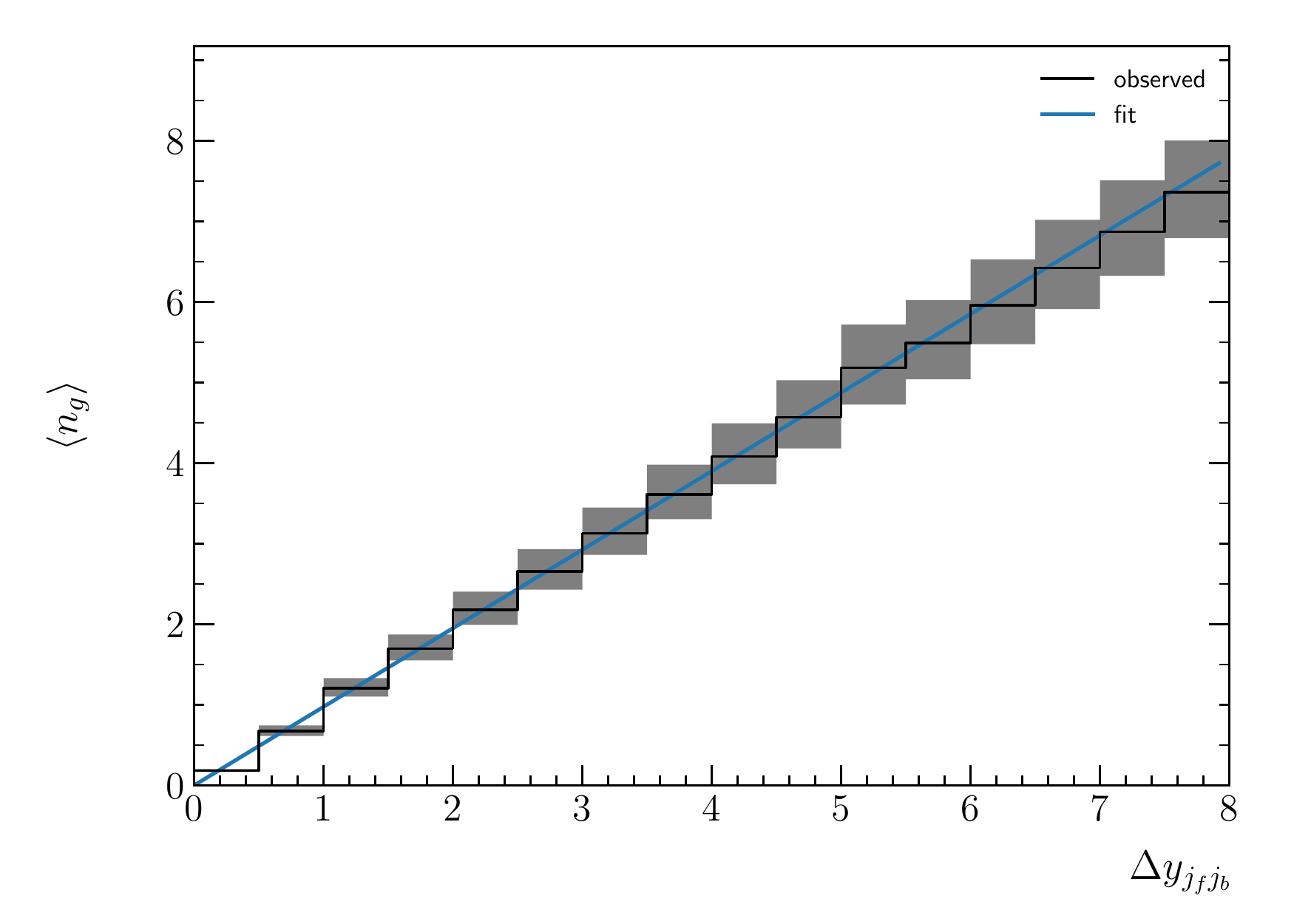}
   \caption{
     Average number of additional gluon emissions $n_g$ as a function of
     the rapidity span between the extremal jets. The bin entries show the
     observed numbers in the production of a Higgs boson in association with
     at least two jets. The solid line shows the function used for the phase
     space generation.
   }
   \label{fig:ng_y}
\end{figure}

\begin{figure}
   \centering
   \includegraphics[width=.49\textwidth]{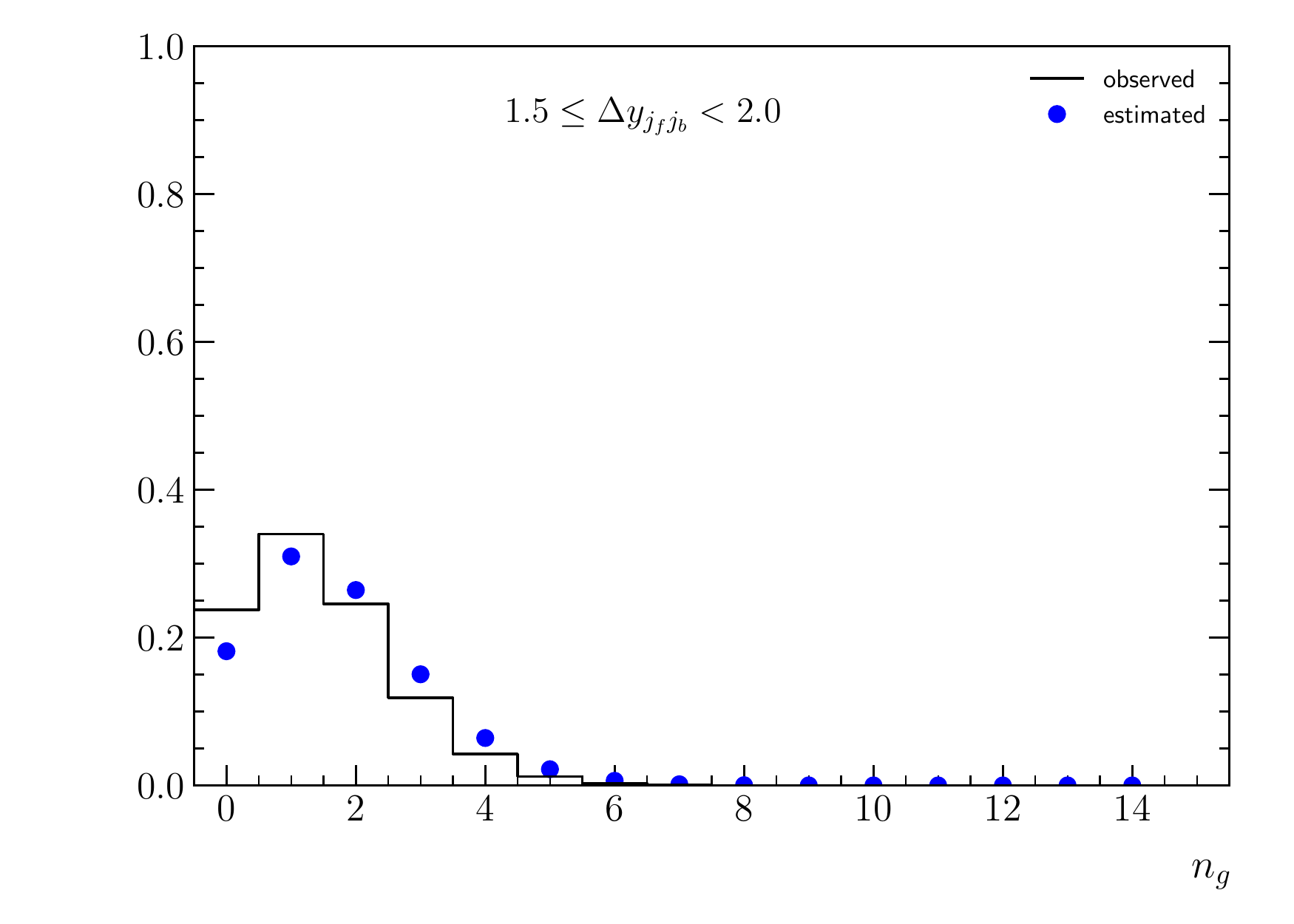}\hfill
   \includegraphics[width=.49\textwidth]{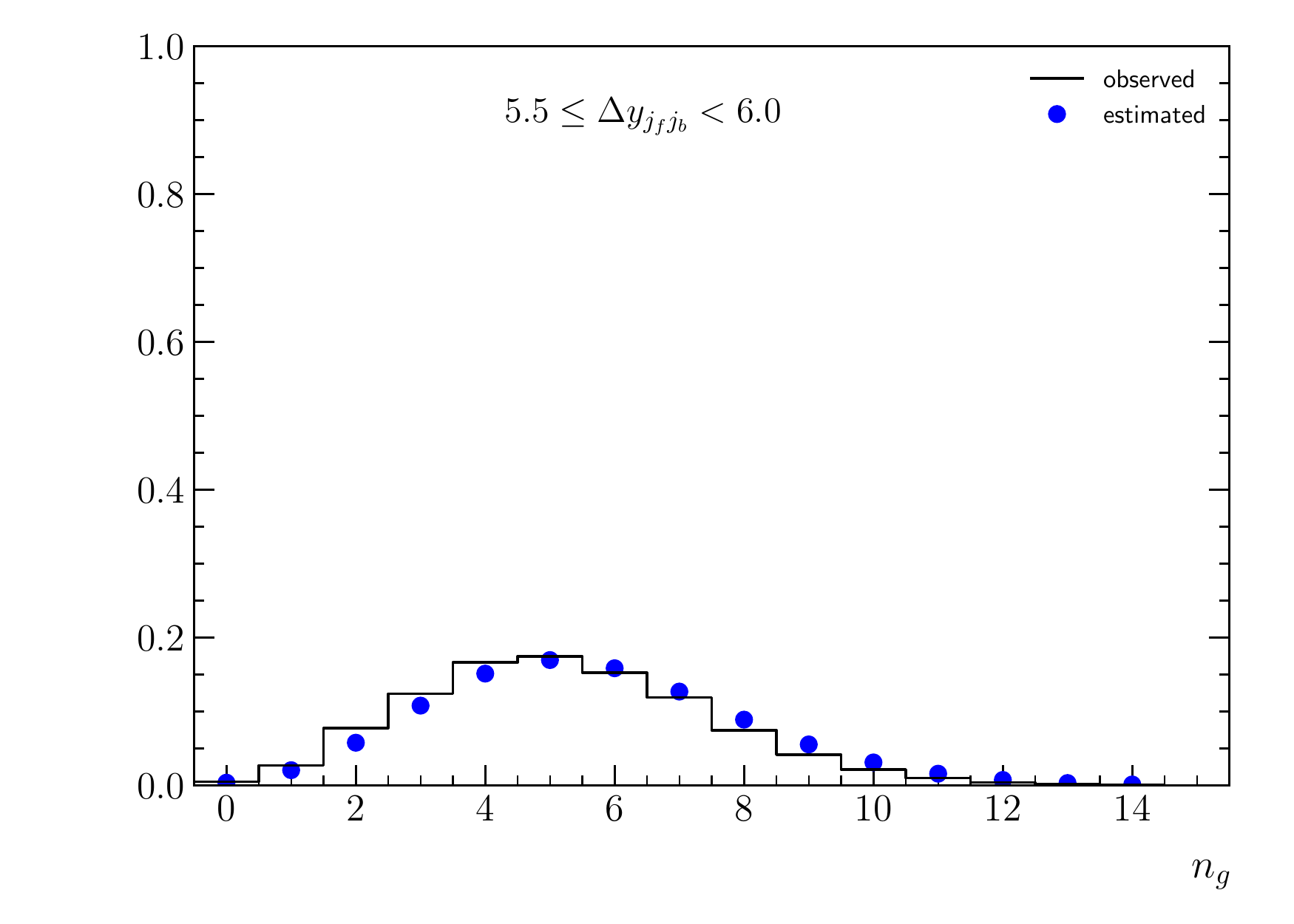}
   \caption{
     Number of additional gluon emissions $n_g$ for two different
     rapidity spans between the extremal jets. The estimates are based on
     Poisson distributions with mean values taken from the fit function in
     Fig.~\ref{fig:ng_y} for $\Delta y_{j_f j_b} = 1.75$ and $\Delta
     y_{j_f j_b} = 5.75$.
   }
   \label{fig:ng_y_poisson}
\end{figure}

For the actual phase space sampling, we assume a Poisson distribution
and extract the mean number of gluon emissions in different rapidity
bins and fit the results to a linear function in $\Delta y_{j_f j_b}$,
finding a coefficient of $0.975$ for the inclusive production of a Higgs
boson with two jets. In Figs.~\ref{fig:ng_y} and \ref{fig:ng_y_poisson}
we compare the fit with the actual outcome.

\subsubsection{Number of Gluons inside Jets}
\label{sec:psp_ng_jet}

For each of the $n_g$ gluon emissions we can split the phase-space
integral into a (disconnected) region inside the jets and a remainder:
\begin{equation}
  \label{eq:psp_split}
  \int [\mathrm{d}p_i] = \int [\mathrm{d}p_i]\,
  \theta\bigg(\sum_{j=1}^{m}{\cal J}_j(p_i)\bigg) + \int [\mathrm{d}p_i]\,
  \bigg[1-\theta\bigg(\sum_{j=1}^{m}{\cal J}_j(p_i)\bigg)\bigg]\,.
\end{equation}
We choose an importance sampling which is flat in the plane spanned by
the azimuthal angle $\phi$ and the rapidity $y$. This is observed
in BFKL and valid in the limit of Multi-Regge-Kinematics (MRK). Furthermore, we assume anti-$k_t$ jets, which cover an area of
$\pi R^2$~\cite{Cacciari:2008gp}.

In principle, the total accessible area in the $y$-$\phi$ plane is given
by $2\pi \Delta y_{fb}$, where $\Delta y_{fb}\geq \Delta y_{j_f j_b}$ is
the a priori unknown rapidity separation between the most forward and
backward partons. In most cases the extremal jets consist of single
partons, so that $\Delta y_{fb} = \Delta y_{j_f j_b}$. For the less common
case of two partons forming a jet we observe a maximum distance of $R$
between the constituents and the jet centre. In rare cases jets have
more than two constituents. Empirically, they are always within a
distance of $\tfrac{5}{3}R$ to the centre of the jet~\cite{Salam}, so
$\Delta y_{fb} \leq \Delta y_{j_f j_b} + \tfrac{10}{3} R$. In practice, the
extremal partons are required to carry a large fraction of the jet
transverse momentum (cf. section~\ref{sec:matching}) and will therefore
be much closer to the jet axis.

\begin{figure}
  \centering
  \includegraphics{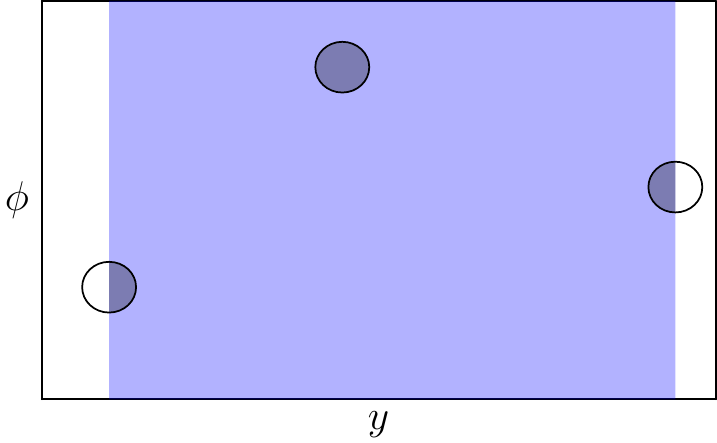} \hfill
  \includegraphics{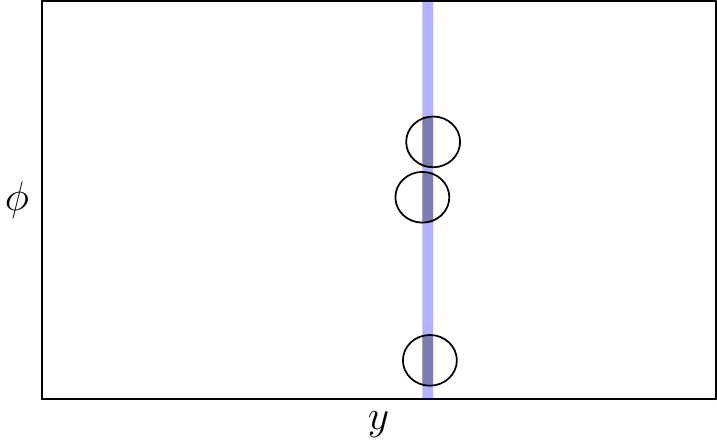}
  \caption{
    Estimated phase space areas for the emission of extra gluons for
    sample three-jet configurations. The left panel shows the case
    of a large rapidity separation. On the right we illustrate the estimate
    for a very small rapidity span.
  }
  \label{fig:ng_ps}
\end{figure}

\begin{figure}
  \centering
  \includegraphics[width=0.75\linewidth]{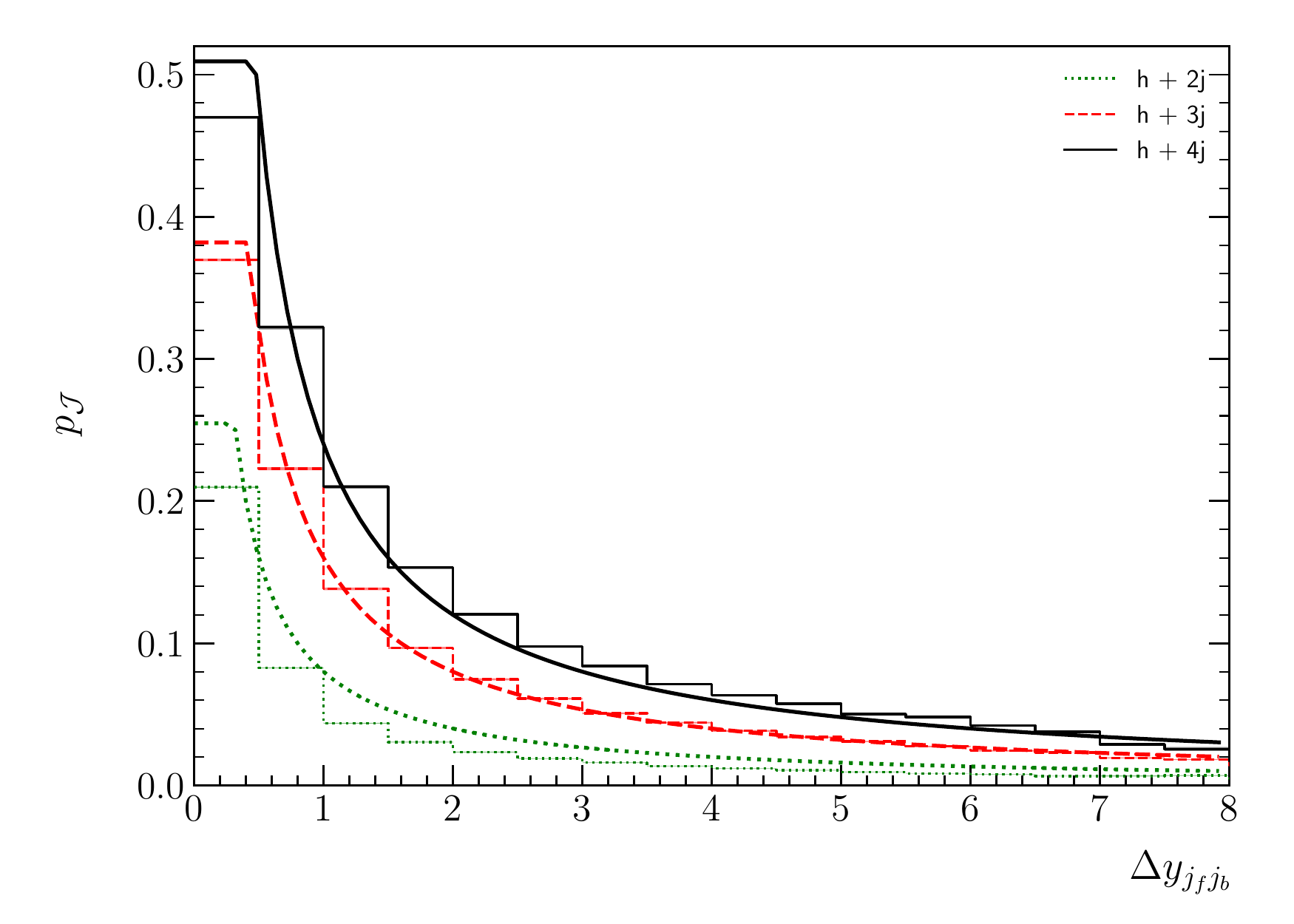}
  \caption{
    Estimated probability for an extra emission to end up inside a jet
    compared to the fraction observed in the exclusive production of a
    Higgs boson with two (green dotted line), three (red dashed line), and four (black
    solid line) jets.
  }
  \label{fig:pJ}
\end{figure}

In summary, for sufficiently large rapidity separations we can use the
approximation $\Delta y_{fb} \approx \Delta y_{j_f j_b}$. If there is no
overlap between jets, the probability $p_{\cal J, >}$ for an extra gluon
to end up inside a jet is then given by (cf. Fig.~\ref{fig:ng_ps})
\begin{equation}
  \label{eq:p_J_large}
  p_{\cal J, >} = \frac{(m - 1)\*R^2}{2\Delta y_{j_f j_b}}\,.
\end{equation}
For a very small rapidity separation, Eq.~\eqref{eq:p_J_large}
obviously overestimates the true probability. The maximum phase space
covered by jets in the limit of a vanishing rapidity distance between
all partons is $2mR \Delta y_{fb}$. We therefore estimate the probability for a parton
to end up inside a jet as
\begin{equation}
  \label{eq:p_J}
  p_{\cal J} = \min\bigg(\frac{(m - 1)\*R^2}{2\Delta y_{j_f j_b}}, \frac{mR}{\pi}\bigg)\,.
\end{equation}
In Fig.~\ref{fig:pJ} we compare this estimate with the actually observed
fraction of additional emissions into jets. We observe good agreement
over the whole rapidity range and for different jet multiplicities.

\subsubsection{Gluons outside Jets and Observed Jet Momenta}
\label{sec:gluons_nonjet}

Using our estimate for the probability of a gluon to be a jet
constituent, we arrive at a number $n_{g,{\cal J}}$ of gluons inside
jets. Before integrating over their remaining phase space, we first have
to determine the momenta $p_{{\cal J}_i}$ of the observed (resummation)
jets from Eq.~\eqref{eq:ptreassign}. To this end, we have to determine
the total transverse momentum $\mathbf{q}_\perp$ of the gluons
\emph{outside} jets. After generating soft transverse momenta
for these
$n_g - n_{g,{\cal J}}$ gluons, we solve the nonlinear system
Eq.~\eqref{eq:ptreassign} using GSL routines~\cite{gsl-ref}. Note
that we have to postpone the rapidity integration, since
at this point the rapidity span in the phase space
integral is not yet known. The most forward and
backward partons have to be part of the extremal jets. Therefore, their momenta will only be determined in the next step.

\subsubsection{Gluons inside Jets}
\label{sec:gluons_jet}

Recall that after the first step in the phase space parametrisation, Eq.~\eqref{eq:select_jets},
each jet has exactly one constituent. We now assign each of the
$n_{g,{\cal J}}$ gluons to a random jet. For jets with a single
constituent, the parton momentum is fixed completely by the constraints
in Eq.~\eqref{eq:resumdijetFKLmatched2}. In the case of two
constituents, we observe that the partons are
always inside the jet cone with radius $R$ and often very close to the
jet centre. This allows an efficient integration by choosing a distance
to the jet centre and an azimuthal angle with respect to the jet axis for
one of the partons, which determines all momentum components of both
constituents.

As is evident from Fig.~\ref{fig:pJ}, jets with three or more
constituents are rare and an efficient phase-space sampling is less
important. For such jets, we exploit the observation that partons with a
distance larger than $R_{\text{max}} = \tfrac{5}{3} R$~\cite{Salam} to
the jet centre are never clustered into the jet. Assuming $N$
constituents, we choose distances, angles, and transverse momenta for
$N-1$ of them and determine the momentum of the last constituent from
the requirement that the constituent momenta have to add up to the jet
momentum. Since this last momentum may lie outside the jet cone, it is
mandatory to check explicitly whether all candidates are actually
clustered into the considered jet. This is to ensure the correct coverage of
phase space.

After constructing the resummation jets, we are now in the position to
evaluate the rapidity integrals for the partons outside the
jets. Finally, we use \texttt{fastjet}~\cite{Cacciari:2011ma} to
recluster all emitted partons into jets again to check whether the
reshuffling conditions imposed by Eq.~\eqref{eq:resumdijetFKLmatched2}
are fulfilled. We also ensure that all partons are assigned as intended,
i.e. the $n_{g,{\cal J}}$ designated jet constituents are indeed part of
their respective jet and all remaining partons end up outside jets.

We have now outlined the practical steps necessary to implement the rewritten
formula of Eq.~\eqref{eq:resumdijetFKLmatched2}.  In the following section we
discuss the results obtained with the new formalism in the key process of Higgs
boson production in association with at least two jets.  Firstly we confirm that if
we limit ourselves to matching with fixed order samples with up to three jets
that we reproduce the results obtained with the previous formalism, but now
with a much higher efficiency.  We will then show and discuss the impact of now
being able to increase the multiplicity in the fixed order samples and also
compare our results to fixed-order next-to-leading order predictions.


\section{Results}
\label{sec:results}

The matching procedure described in this work is significantly more
efficient and flexible than the approach used in previous versions of
\HEJ. To illustrate this, we present new results for the production of a
Higgs boson in association with at least two jets matched to
leading-order events with up to four jets. Previously, matching of \HEJ
to just three jets was achieved for this process, while using significantly more
CPU resources than necessary with the current approach. In its new
formulation, the matching is in practice only limited by the
capabilities of the underlying fixed-order generator. For instance, the
generation of one set of $1000$ unweighted leading-order events for
the production of a Higgs boson with four jets typically took a few CPU days
using \madgraph\cite{Alwall:2014hca}. It is then just a few additional
CPU seconds to generate $100$ resummation events from each of the
fixed-order 4-jet events, so $100\,000$ trial resummation configurations
in total. In the previous matching approach, generating $100\,000$
resummation configurations would require the same number of fixed-order
matrix element evaluations. In this sense the new matching algorithm is
more efficient by a factor of almost 100.\footnote{Note, however, that
  for our specific setup a direct comparison between both approaches is
  not possible since there is no analogue to the concept of unweighted
  fixed-order events in the ``old'' approach.}

This section will present the results obtained with the new procedure for
matching and resummation. Section~\ref{sec:setup} describes the cuts and
analysis used. Section~\ref{sec:comp} compares new results with matching up
to three jets with those obtained previously, and demonstrates that the two
methods yield equivalent results. Section~\ref{sec:res_4j} investigates the
stability of the results obtained by investigating the impact of
increasing the order to which matching is achieved. In general, the matching
to higher multiplicities should have little impact for configurations where
the four-jet contribution is insignificant or the approximation within \HEJ already
provides a good description. Conversely, the corrections from matching to
successive multiplicities can serve to indicate the stability of the \HEJ
predictions for observables sensitive to additional hard radiation. Finally,
in Section~\ref{sec:NLOcomp} we match the inclusive $Hjj$-cross section to
NLO accuracy, thus obtaining the most precise predictions for
$Hjj$-production, including the effects of VBF cuts and central jet
vetos. These results are compared to those obtained at fixed next-to-leading
order accuracy.

\subsection{Setup}
\label{sec:setup}
To facilitate the comparison with previous results we will adopt the cuts of the
experimental analysis of Ref.~\cite{Aad:2014lwa}, and the parameters of our
analysis in Ref.~\cite{Andersen:2017kfc}. To recapitulate, we consider the
gluon-fusion-induced production of a Higgs boson together with at least two
anti-$k_t$ jets with transverse momenta $p_{\perp, j} > 30\,$GeV, rapidities
$|y_j| < 4.4$, and radii $R = 0.4$ at the $13\,$TeV LHC. While it is obviously
irrelevant for the considerations of the QCD corrections considered in this
paper, we consider the Higgs boson decay into two photons with
\begin{align}
  \label{eq:photon_cuts}
  |y_{\gamma}| &< 2.37, \qquad 105~\textrm{GeV} <
  m_{\gamma_1 \gamma_2} < 160~\textrm{GeV},\notag\\
  p_{\perp, \gamma_1} &> 0.35\,m_{\gamma_1 \gamma_2}, \qquad p_{\perp, \gamma_2} > 0.25\,m_{\gamma_1 \gamma_2},
\end{align}
and separations $\Delta R(\gamma, j), \Delta R(\gamma_1, \gamma_2)  >
0.4$ from the jets and each other. To be consistent with our previous
analysis we set the Higgs-boson mass to $m_H=125\,$GeV, a width of
$\Gamma_H = 4.165\,$MeV and a branching fraction of $0.236\%$ for the
decay into two photons. We use the CT14nlo PDF set~\cite{Dulat:2015mca}
as provided by LHAPDF6~\cite{Buckley:2014ana}.

In addition to inclusive quantities with the basic cuts listed above, we
also consider additional VBF-selection cuts applied to the hardest jets as
in~\cite{Aad:2014lwa}:
\begin{align}
  \label{eq:vbfcuts}
  |y_{j_1}-y_{j_2}| > 2.8, \qquad m_{j_1 j_2} > 400~\textrm{GeV}.
\end{align}

In the first step, we generate leading-order events with two, three and four
jets. With our new matching procedure we are free to use an arbitrary
fixed-order event generator for this purpose. For the present analysis
we employ version 2.5.5 of \madgraph~\cite{Alwall:2014hca}. For each jet
multiplicity we produce about 2000 sets of unweighted events, each
comprising 10\,000 events for the sets with two or three jets and 1000
events for sets with four jets.

As the transverse momenta of the jets are modified during resummation
(cf. Eq.~\eqref{eq:ptreassign}), we have to generate at least a fraction
of events with Born-jet momenta below the threshold of $30\,$GeV
required from the observed jets. As already shown in
Fig.~\ref{fig:pjperpmin} the contribution after the resummation from
such tree-level configurations in the matching drops off very rapidly
below the jet transverse momentum analysis scale of $30\,$GeV. Passing
this information to the underlying fixed-order generator, such that only
a small fraction of events are generated below the nominal transverse
momentum threshold could improve the sampling considerably. Having such
an option would therefore be highly desirable. For the time being, we
manually generate 200 additional sets of Born-level events with
transverse momenta down to $20\,$GeV for each jet multiplicity.

Events with more exclusive jets than can be reasonably
evaluated in \madgraph are unmatched and generated with a custom built
Monte Carlo generator based on tree-level \HEJ matrix elements instead
of full leading-order ones. In this way, we can supplement the
fixed-order input with events including up to ten jets obtained within
the \HEJ approximation. These events are simply passed through the same
matching mechanism based on Eq.~\eqref{eq:resumdijetFKLmatched2} just as
the lower-multiplicity events obtained using \madgraph. The maximum
multiplicity of ten is an arbitrary cut-off, based on an explicit check
that the impact on observables at this multiplicity is negligible.

Since the final kinematics required for a kinematic scale setting are not
known at the point of generating fixed-order events, we use a fixed
renormalisation and factorisation scale of $\mu_r = \mu_f = m_H$ during the
fixed-order generation. After resummation the events are rescaled to a
central scale of $\mu_r = \mu_f = H_T/2$. In order to assess the scale
dependence, we independently vary both the renormalisation and factorisation scales by factors
$\{1/2, 1/\sqrt{2}, 1, \sqrt{2}, 2\}$ and discard combinations with
$\mu_r/\mu_f < 1/2$ or $\mu_r/\mu_f > 2$. In the effective Higgs-gluon coupling, we
keep the renormalisation scale at the Higgs-boson mass and apply the limit of
an infinite top-quark mass. These scale settings and even the use of the
infinite top-mass limit are however not inherent to the use of the high-energy
resummation of \HEJ, but can be included with modified components of the amplitudes, similar to Ref.~\cite{DelDuca:2003ba}.

After generating the tree-level input events, we apply resummation as presented in
the previous sections. Recent progress described
in~\cite{Andersen:2017kfc} allows us to apply resummation not just for
FKL-ordered matching-events, but also the sub-leading contribution from
events with three jets or more, where the rapidity-ordering of the two most forward or
most backward jets is flipped compared to FKL ordering. This corresponds to a
gluon emission outside a rapidity-interval delimited by quark jets. For each resummation-type tree-level event, we generate 100 trial
configurations in the resummation phase space.  For the remaining sub-leading
events we cannot add resummation and
simply adjust the factorisation and renormalisation scales as described
above.

\subsection{Comparison to previous results}
\label{sec:comp}
\begin{figure}
  \centering
  \begin{subfigure}[t]{0.495\textwidth}
    \includegraphics[width=\linewidth]{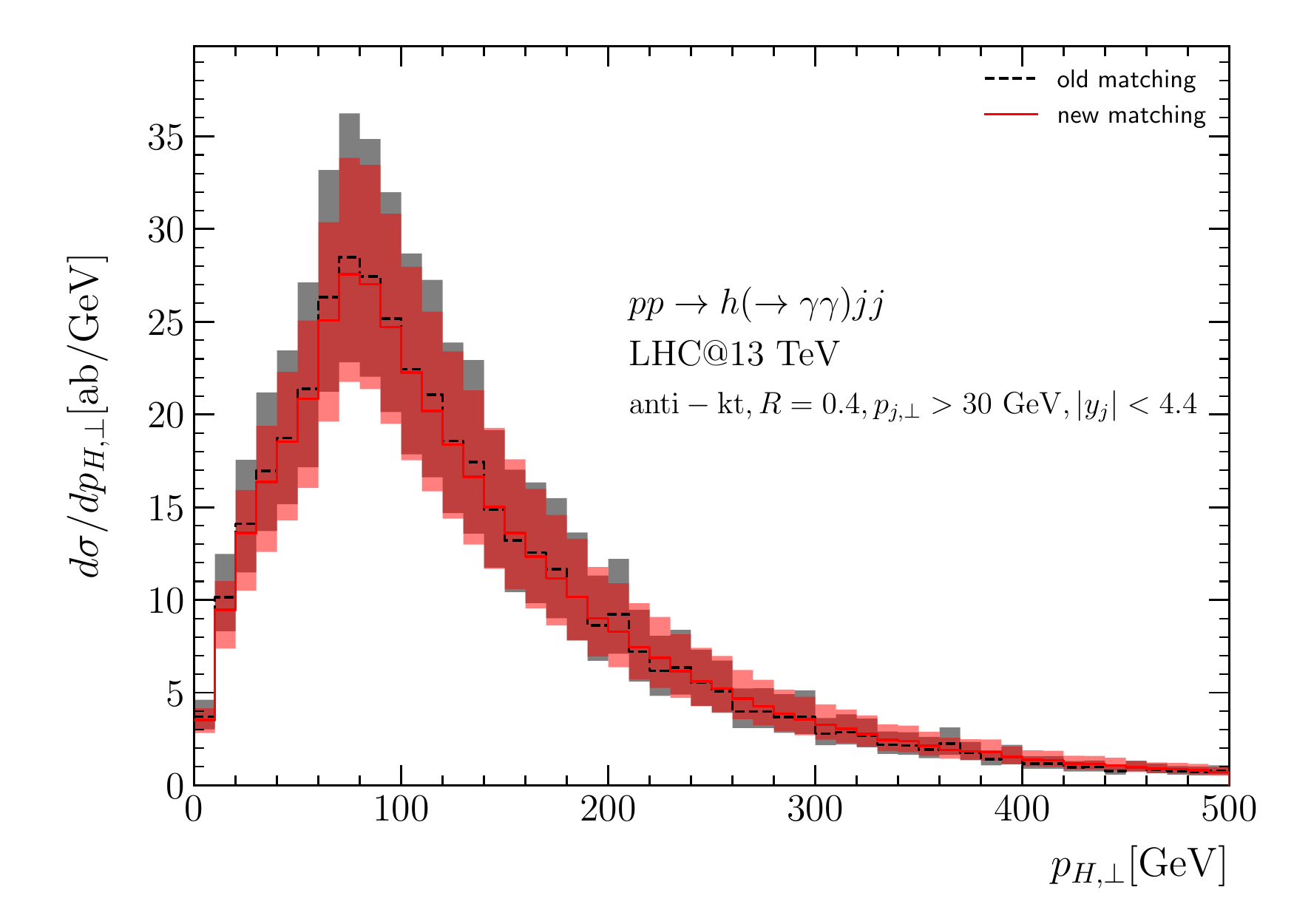}
    \caption{}
  \end{subfigure}
  \begin{subfigure}[t]{0.495\textwidth}
    \includegraphics[width=\linewidth]{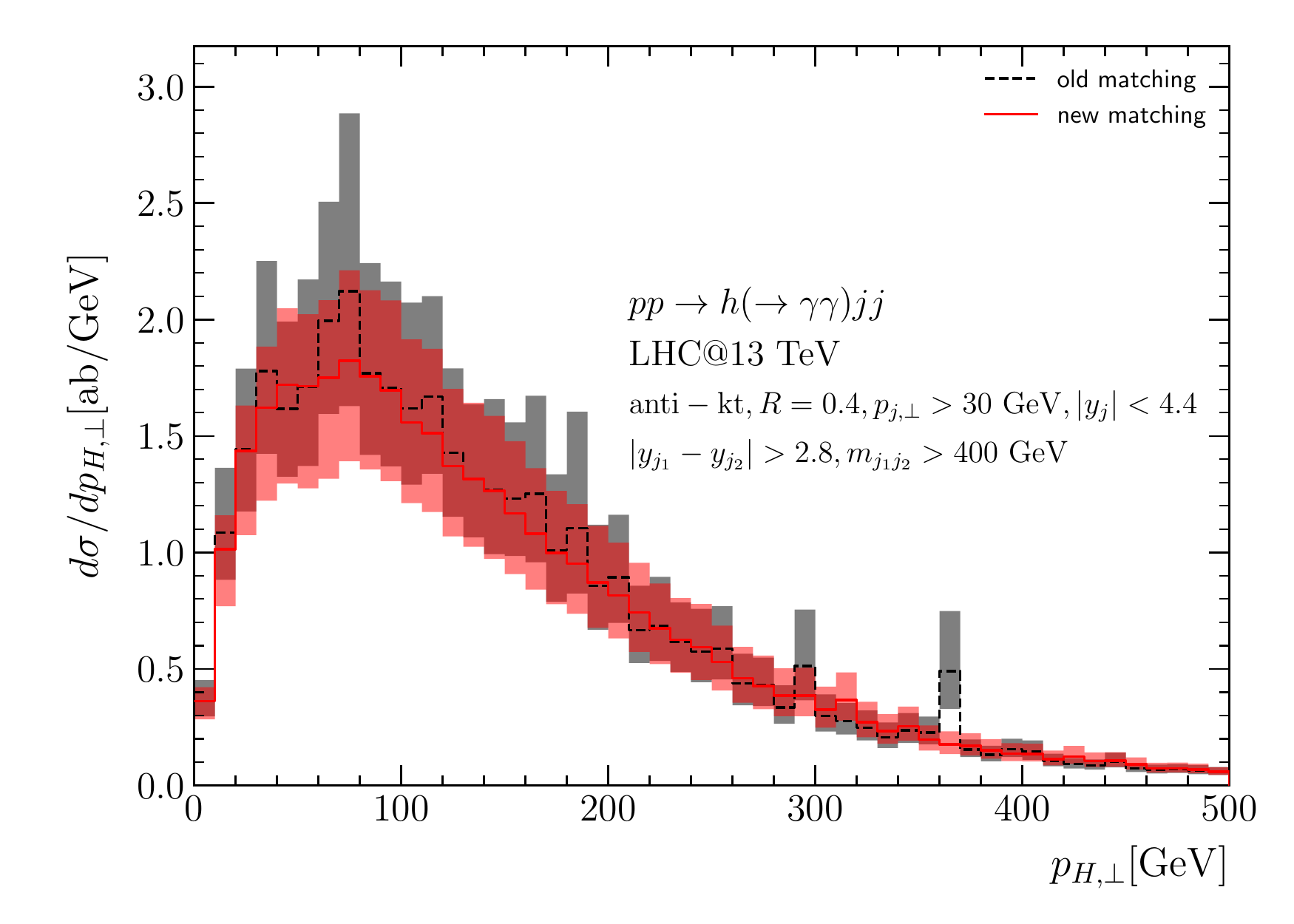}
    \caption{}
  \end{subfigure}
    \caption{
      Comparison of the new matching procedure to previous \HEJ
      results obtained in~\cite{Andersen:2017kfc}. The panels show the
      transverse momentum distributions of the Higgs boson for (a) inclusive
      cuts and (b) VBF cuts.
    }
  \label{fig:HEJ_cmp}
\end{figure}

In order to demonstrate the validity of the new approach we compare here
first our
results with leading-order matching up to three jets to those obtained in our previous work~\cite{Andersen:2017kfc} .  We find good agreement within
the statistical errors. As examples, we show the transverse momentum
distributions for the Higgs boson for inclusive and VBF cuts in
Fig.~\ref{fig:HEJ_cmp}. The previous and new method of organising the
calculation are equivalent. For the
comparison, we have adjusted our settings to match those
in~\cite{Andersen:2017kfc} as closely as possible. Apart from restricting the fixed-order matching to configurations with at most three
jets, this also means that the extremal partons are required to have a fixed minimum transverse
momentum of $27\,$GeV instead of a fraction of the corresponding jet
momentum, as discussed in section~\ref{sec:matching}.

\subsection{Impact of four-jet matching on Distributions}
\label{sec:res_4j}
The \HEJ approximation is exact in the limit of Multi-Regge kinematics,
i.e.~for large rapidity separation between hard jets. An equivalent
characterisation is to demand the centre-of-mass energy and the
invariant masses between all final-state jets to be much larger
than the typical transverse momenta of these. If these conditions are fulfilled, we
expect a good \HEJ prediction and hence small matching corrections. In order
to assess the perturbative stability of the final predictions, we will here
study the impact on the resummed and matched cross section of scale
variations and of successive matching to two-jet, three-jet and four-jet
tree-level events.

\begin{figure}
  \centering
  \begin{subfigure}[t]{0.495\textwidth}
    \includegraphics[width=\linewidth]{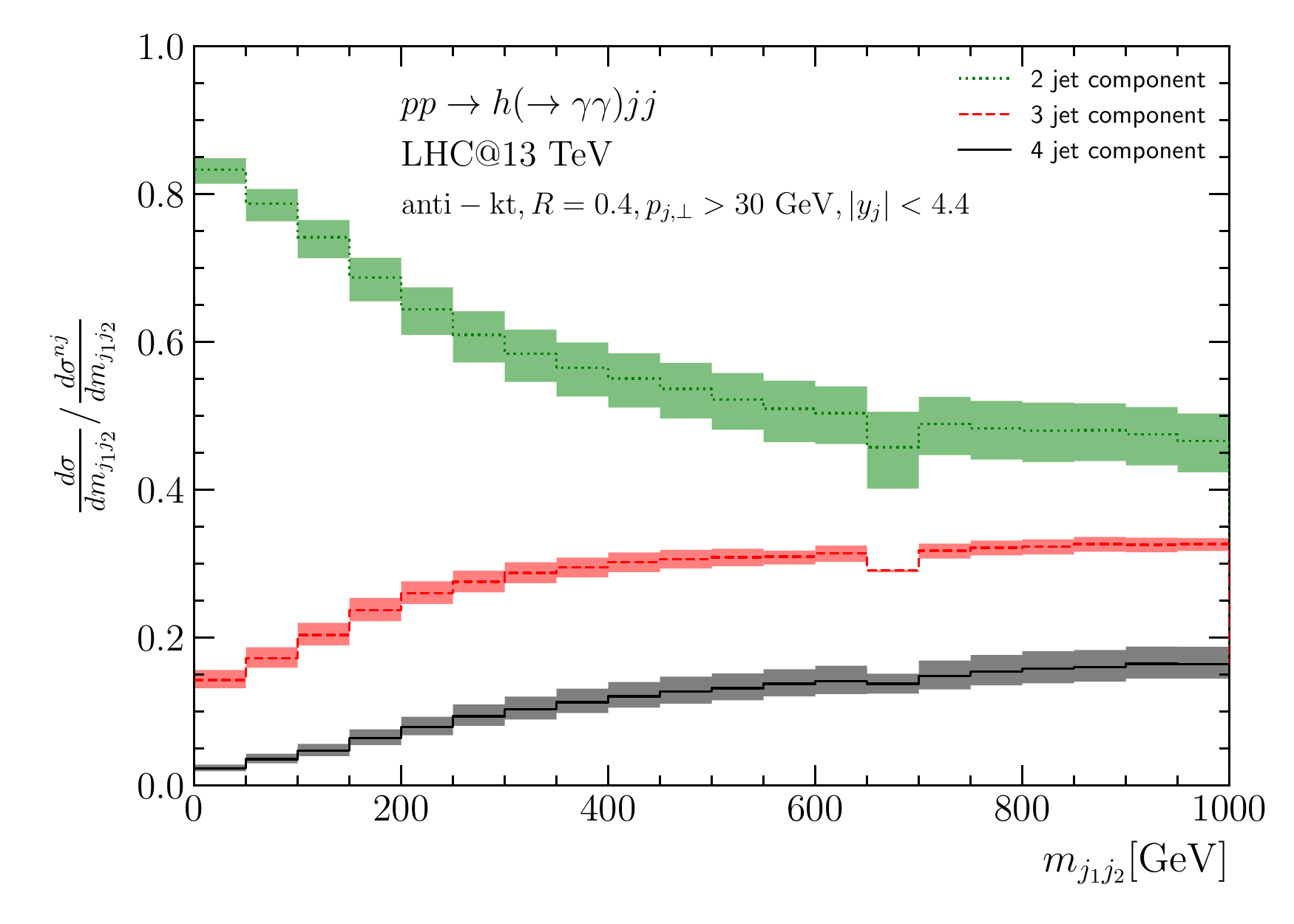}
    \caption{}
  \end{subfigure}
  \begin{subfigure}[t]{0.495\textwidth}
    \includegraphics[width=\linewidth]{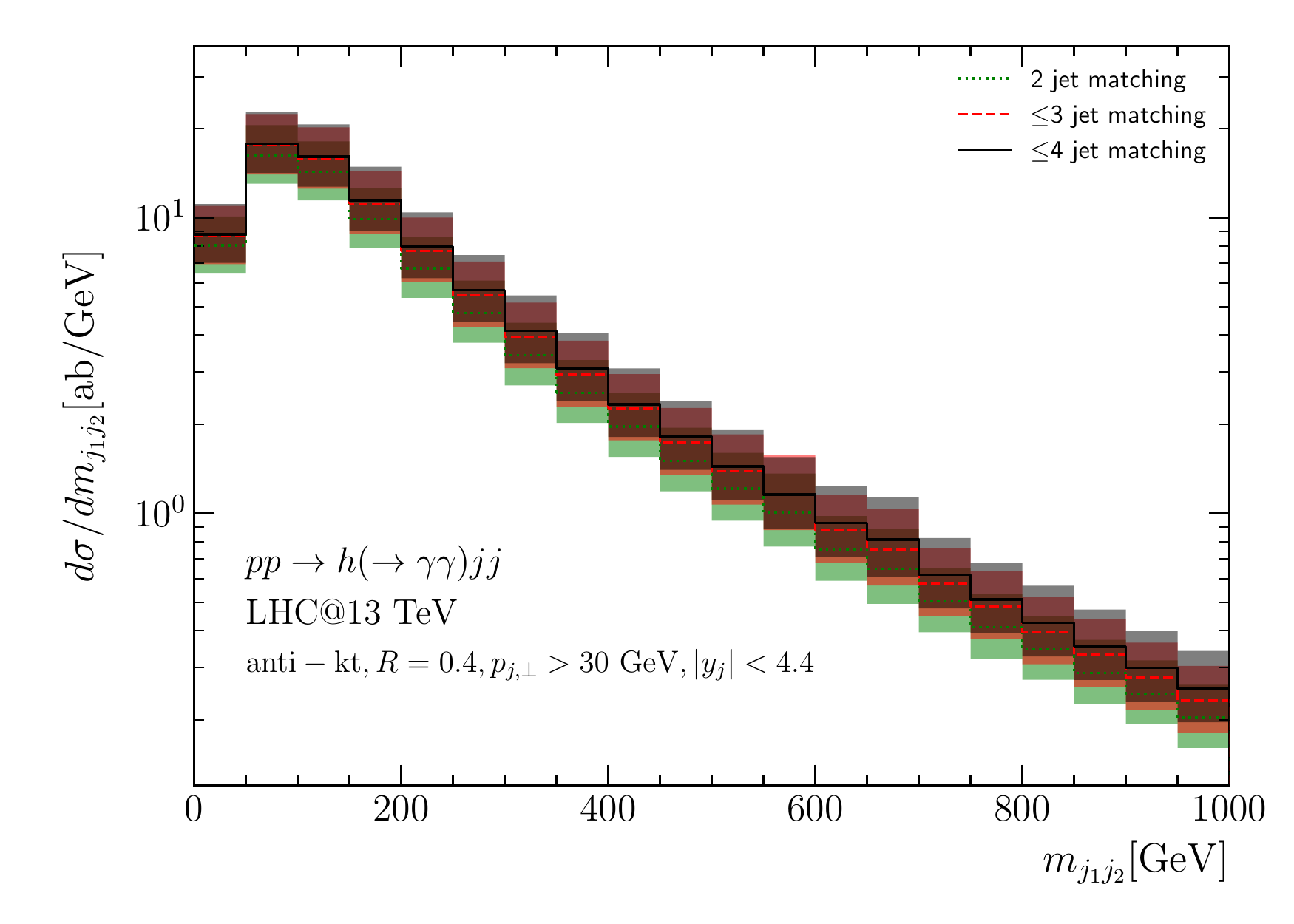}
    \caption{}
  \end{subfigure}
  \caption{Distribution of the invariant mass between the hardest
    jets. Panel (a) shows the fractional contributions from exclusive
    two-, three-, and four-jet events. Panel (b) depicts the effects of
    fixed-order matching up to two, three, and four jets.}
  \label{fig:m12}
\end{figure}

One of the main goals of \HEJ is to improve the prediction of the
gluon-fusion background to Higgs-boson production in weak-boson
fusion. Standard VBF cuts project out a kinematic region with a large
invariant mass between the hardest jets, where the gluon fusion receives
significant contributions from higher jet multiplicities. Fig.~\ref{fig:m12}(a)
displays the relative contribution of the exclusive two-, three- and four-jet
component to the distribution on the invariant mass between the two hardest
(in transverse momentum) jets. The relative contribution from  exclusive three- and
four-jet-events increases with increasing
$m_{{j_1}{j_2}}$. Fig.~\ref{fig:m12}(b) displays the impact of matching of
successive multiplicity on the distribution of the invariant mass between the
two hardest (in transverse momentum) jets. The effect of
the four-jet matching is small but non-zero even at large $m_{j_1 j_2}$. This
is because even in this limit a large separation between \emph{all} jets is
not guaranteed.

The contribution from jet multiplicities of more than or equal to 5 is less than $5\%$ for an invariant mass
of at least $1\,$TeV. We conclude that the uncertainty on the distribution of
$m_{{j_1}{j_2}}$ from terminating the matching at the four-jet contribution
is insignificant, and well within the quoted scale variation.

\begin{figure}
  \centering
  \begin{subfigure}[t]{0.495\textwidth}
    \includegraphics[width=\linewidth]{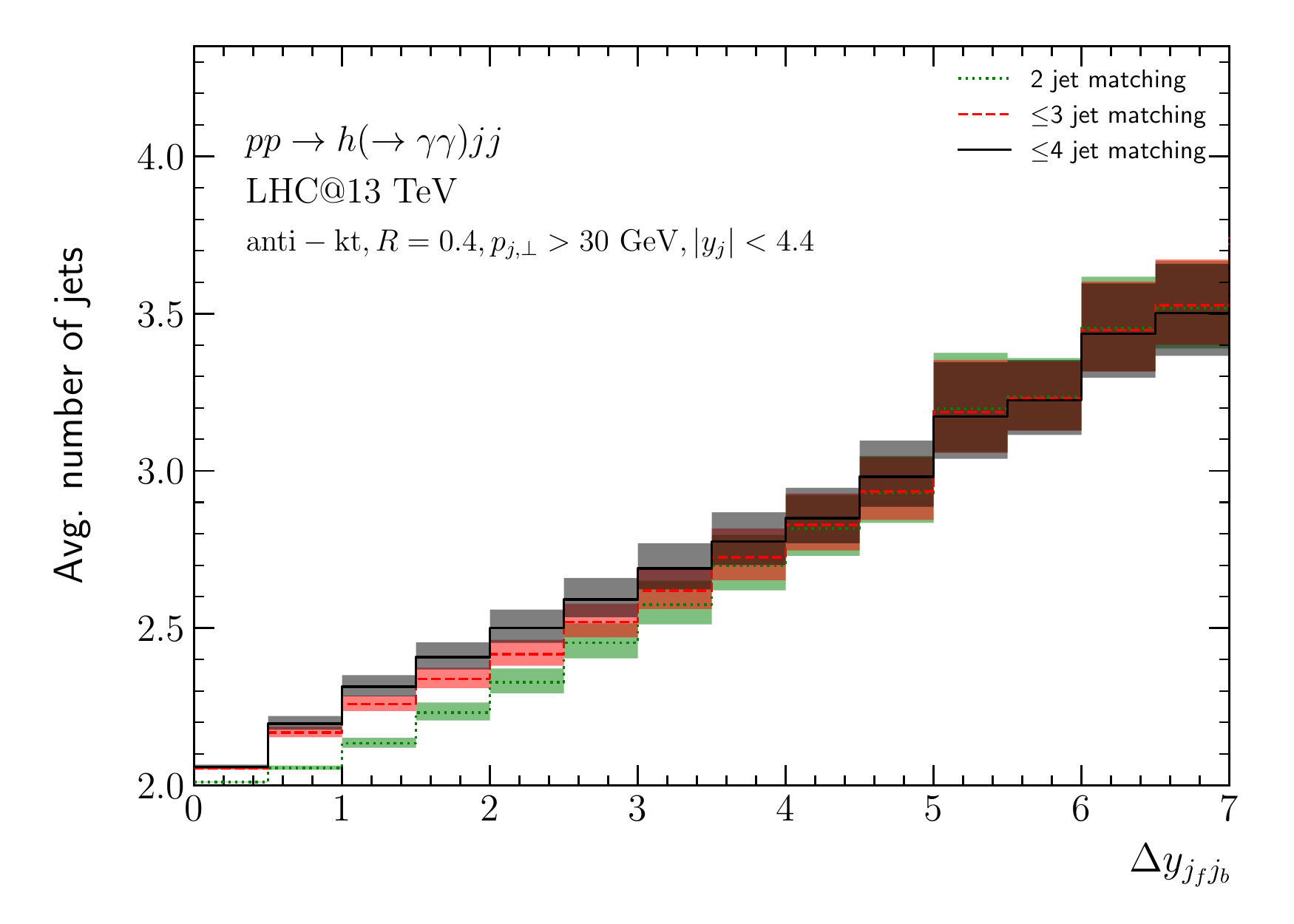}
  \caption{}
  \end{subfigure}
  \begin{subfigure}[t]{0.495\textwidth}
    \includegraphics[width=\linewidth]{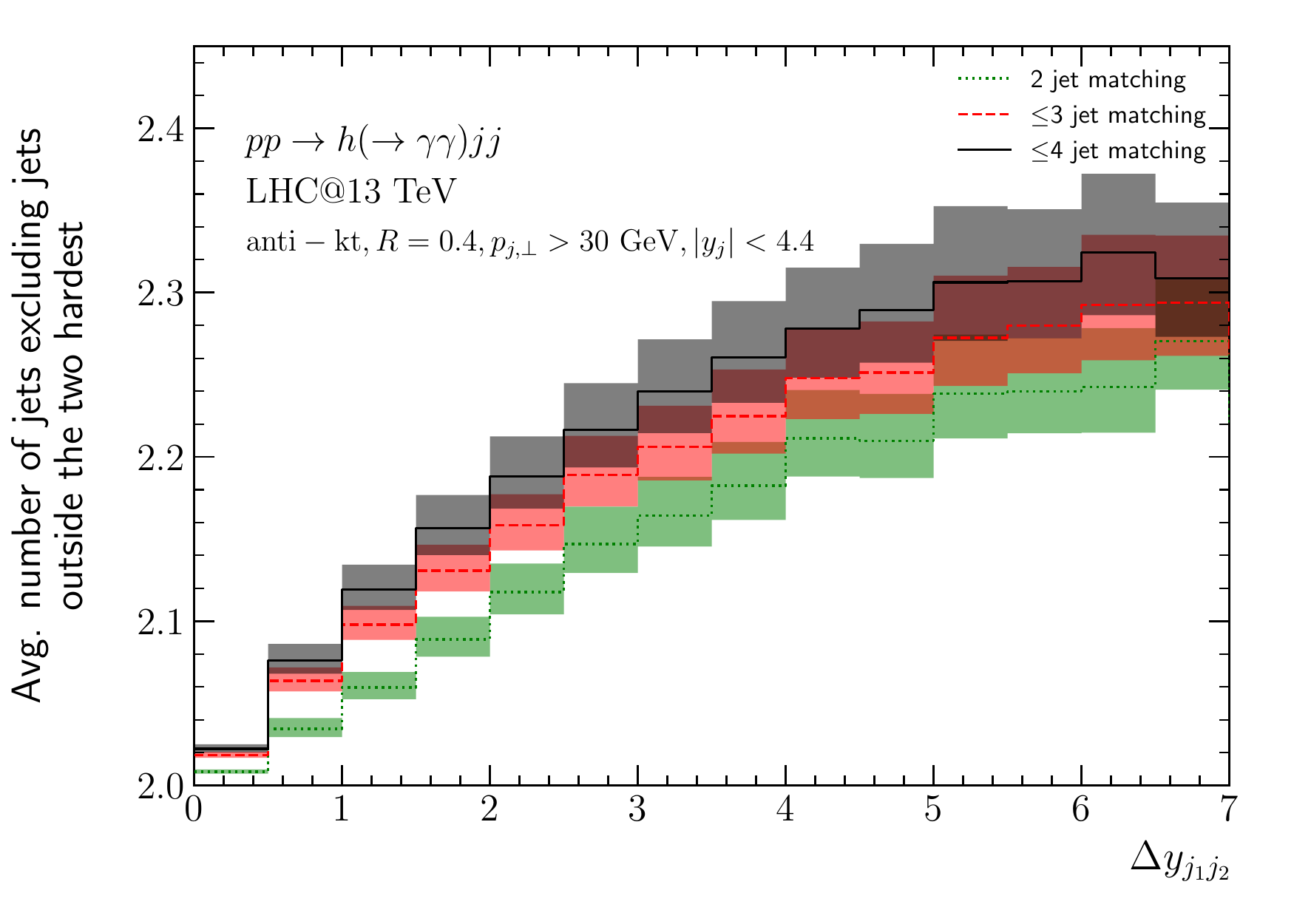}
    \caption{}
  \end{subfigure}
  \caption{Average number of jets for fixed-order matching up to two,
    three, and four jets. In (a) we show the average total number of
    jets vs.~the maximum rapidity-separation. In (b) we show the number of
    jets in the rapidity region of the two hardest jets.}
  \label{fig:njets}
\end{figure}

A central prediction of BFKL, which arises also within \HEJ, is a linear
increase in the number of jets for a growing rapidity span between the most
backward and forward jets.\footnote{This growth continues until the invariant
  mass of just the forward and backward jets is so large that no other jets
  can be emitted due to energy and momentum constraints. The fixed-oder NLO
  results have a similar behaviour at small $\Delta y_{j_f,j_b}$ until the
  average jet multiplicity is saturated by the fixed-order truncation.} This behaviour is
demonstrated in Fig.~\ref{fig:njets}, which also investigates the impact of
matching to tree-level of successive multiplicities. Although the contribution from higher
jet multiplicities increases with the rapidity separation, the effect of
fixed-order matching on this observable actually decreases. This confirms our
expectation that the \HEJ approximation works well for large $\Delta
y_{j_f,j_b}$. It is this linear increase in the average number of jets
versus increasing rapidity span which can be exploited to suppress the
gluon-fusion contribution with a central jet veto.

In contrast to this, if the two \emph{hardest} jets are tagged, and only jets
in-between these are counted as a function of the rapidity difference between
the hardest jets, then the initial linear growth stalls at an average number
of jets of around 2.3. The difference in behaviour
to the VBF contribution is therefore less pronounced by tagging the hardest
jets, rather than the most forward and backward hard jets. This was
investigated further in Ref.\cite{Bendavid:2018nar}. Also, the impact of the
matching corrections remains sizeable for all rapidity separations.

\begin{figure}
  \centering
  \begin{subfigure}[t]{0.495\textwidth}
  \includegraphics[width=\linewidth]{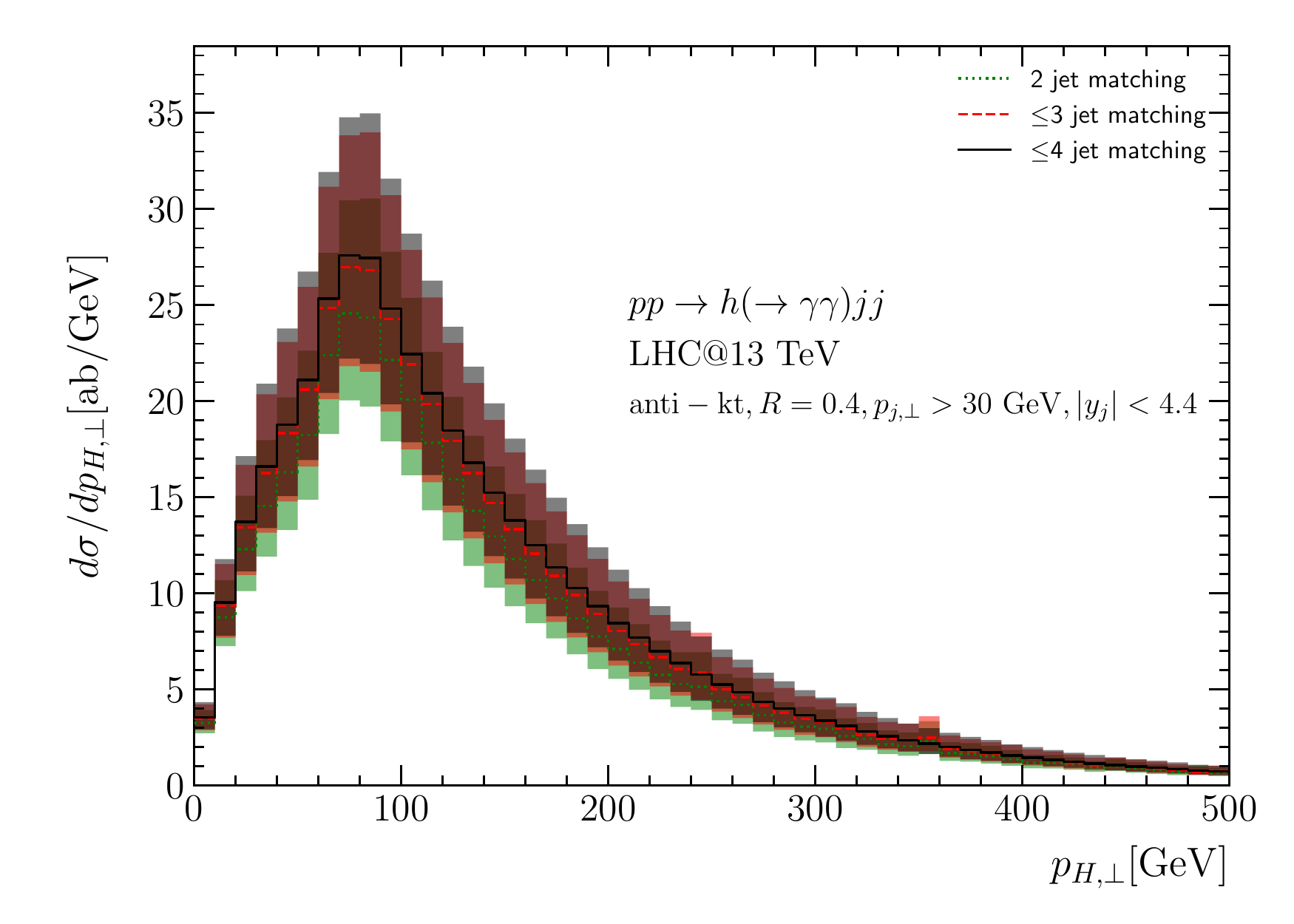}
  \caption{}
  \end{subfigure}
  \begin{subfigure}[t]{0.495\textwidth}
    \includegraphics[width=\linewidth]{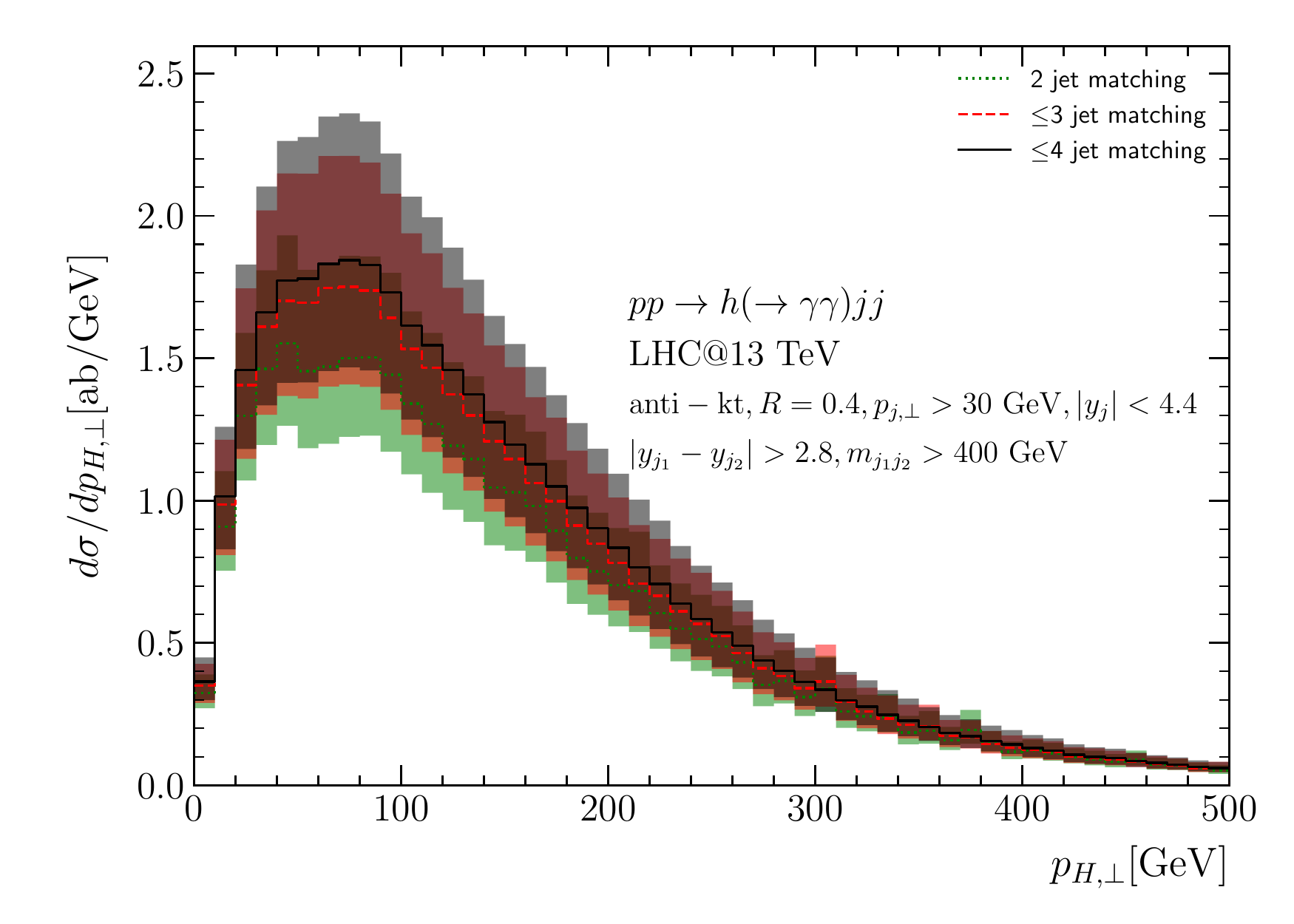}
    \caption{}
  \end{subfigure}
  \caption{Distribution of the Higgs-boson transverse momentum with (a)
    inclusive and (b) VBF cuts.}
  \label{fig:ptH}
\end{figure}

In observables which are neither dominated by higher jet multiplicities
nor completely described by the \HEJ approximation we observe that the
matching corrections are converging, but the corrections from four-jet
matching are non-negligible. In
Fig.~\ref{fig:ptH} we show the distribution of the Higgs-boson
transverse momentum with inclusive and with VBF cuts. While there is a
notable difference between the matching to fixed-order predictions up to
two and three jets, the effect of four-jet matching is much smaller. In
all cases the matching corrections are well inside the scale variation.

\begin{figure}
  \centering
  \begin{subfigure}[t]{0.495\textwidth}
  \includegraphics[width=\linewidth]{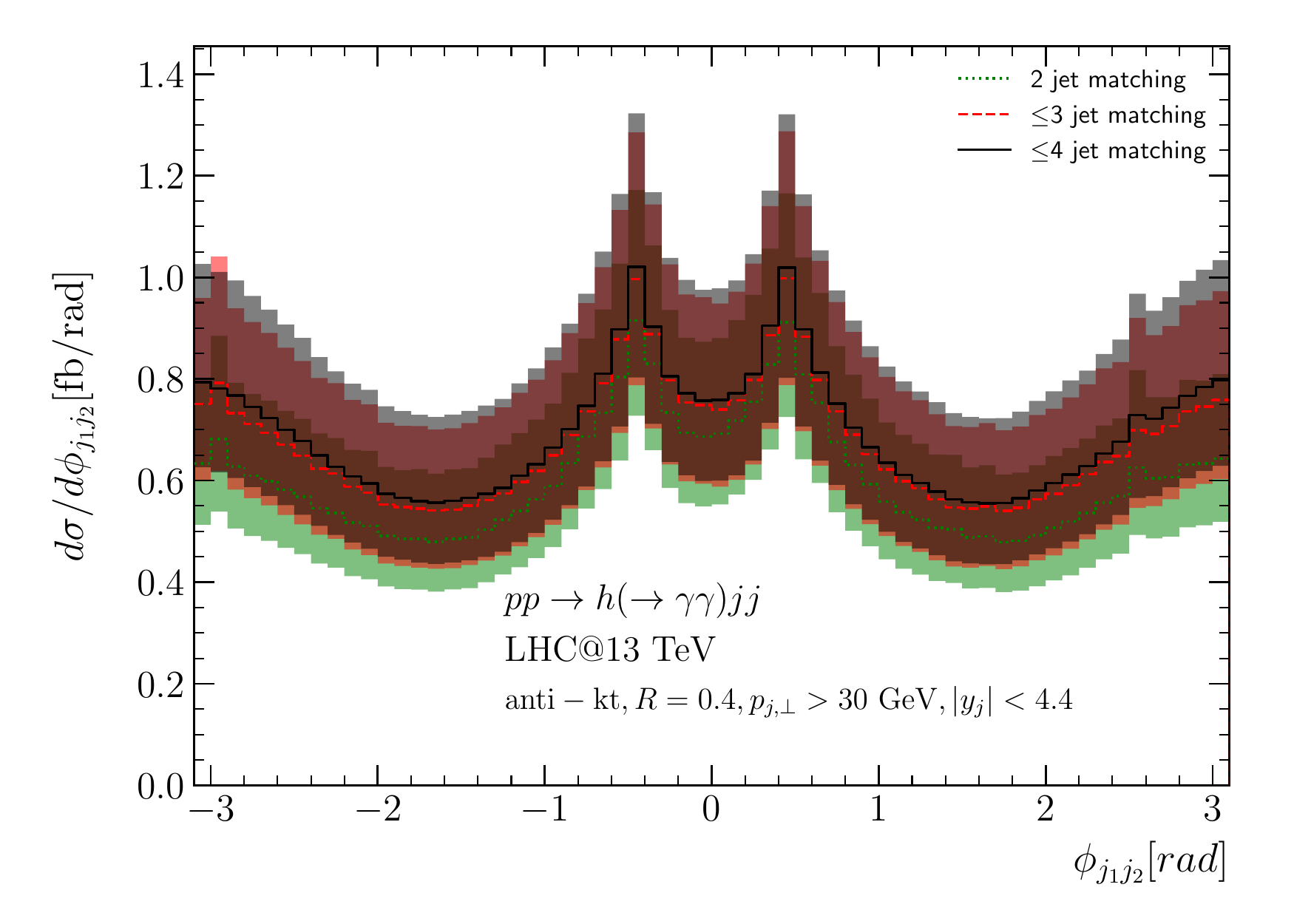}
  \caption{}
  \end{subfigure}
  \begin{subfigure}[t]{0.495\textwidth}
    \includegraphics[width=\linewidth]{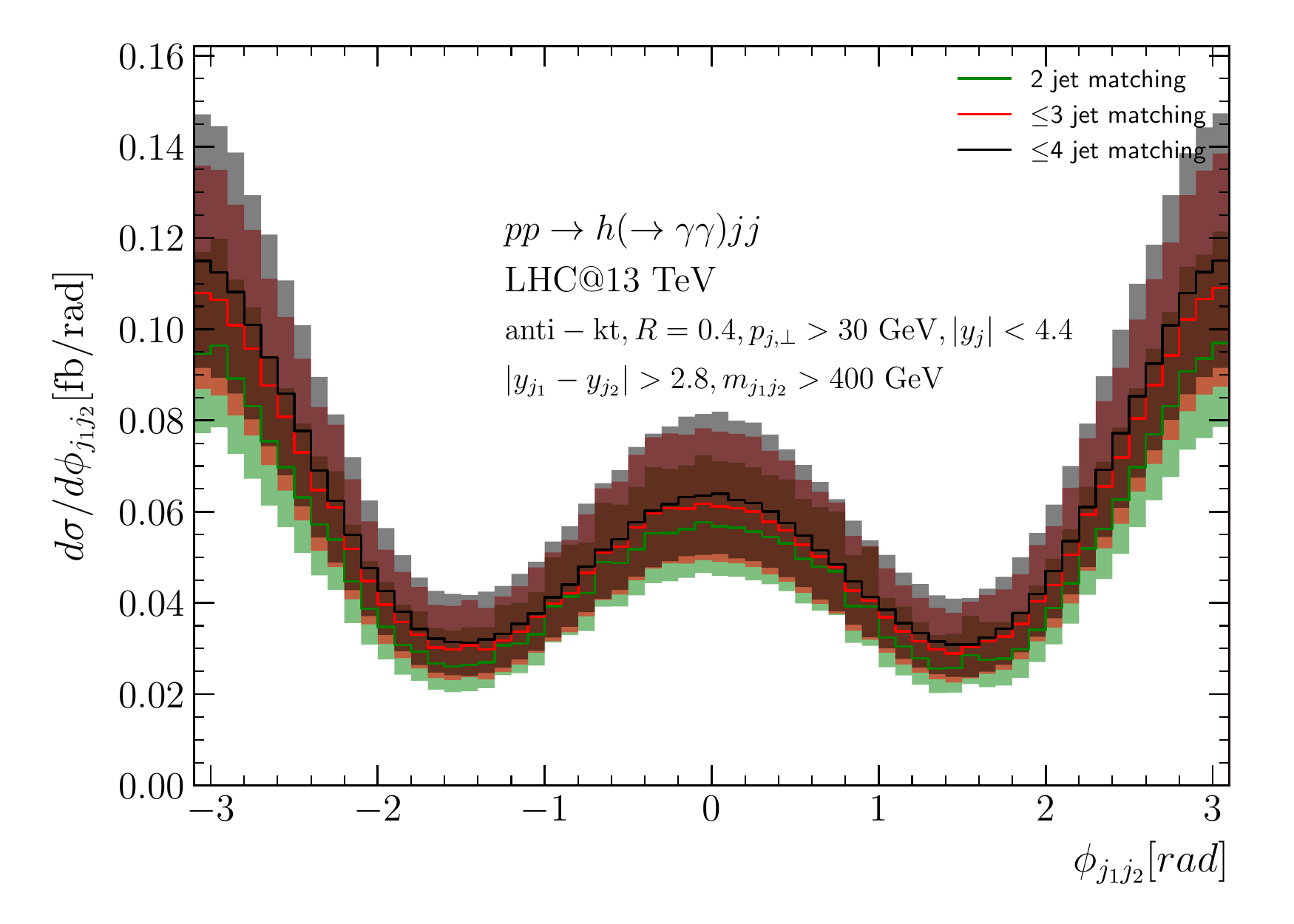}
    \caption{}
  \end{subfigure}
  \caption{Distribution of the azimuthal angle between the two hardest
    jets with (a) inclusive and (b) VBF cuts.}
  \label{fig:phi12}
\end{figure}

The azimuthal angle between jets is of particular interest for the
extraction of the $CP$-properties of the effective coupling between the
Higgs boson and gluons. Fig.~\ref{fig:phi12} shows the effects of
fixed-order matching on the distribution of the angle between the two
hardest jets. Similar to the transverse momentum distribution in
Fig.~\ref{fig:ptH} the corrections from four-jet matching are uniformly
moderate.

\begin{figure}[tb]
  \centering
  \begin{subfigure}[t]{0.495\textwidth}
  \includegraphics[width=\linewidth]{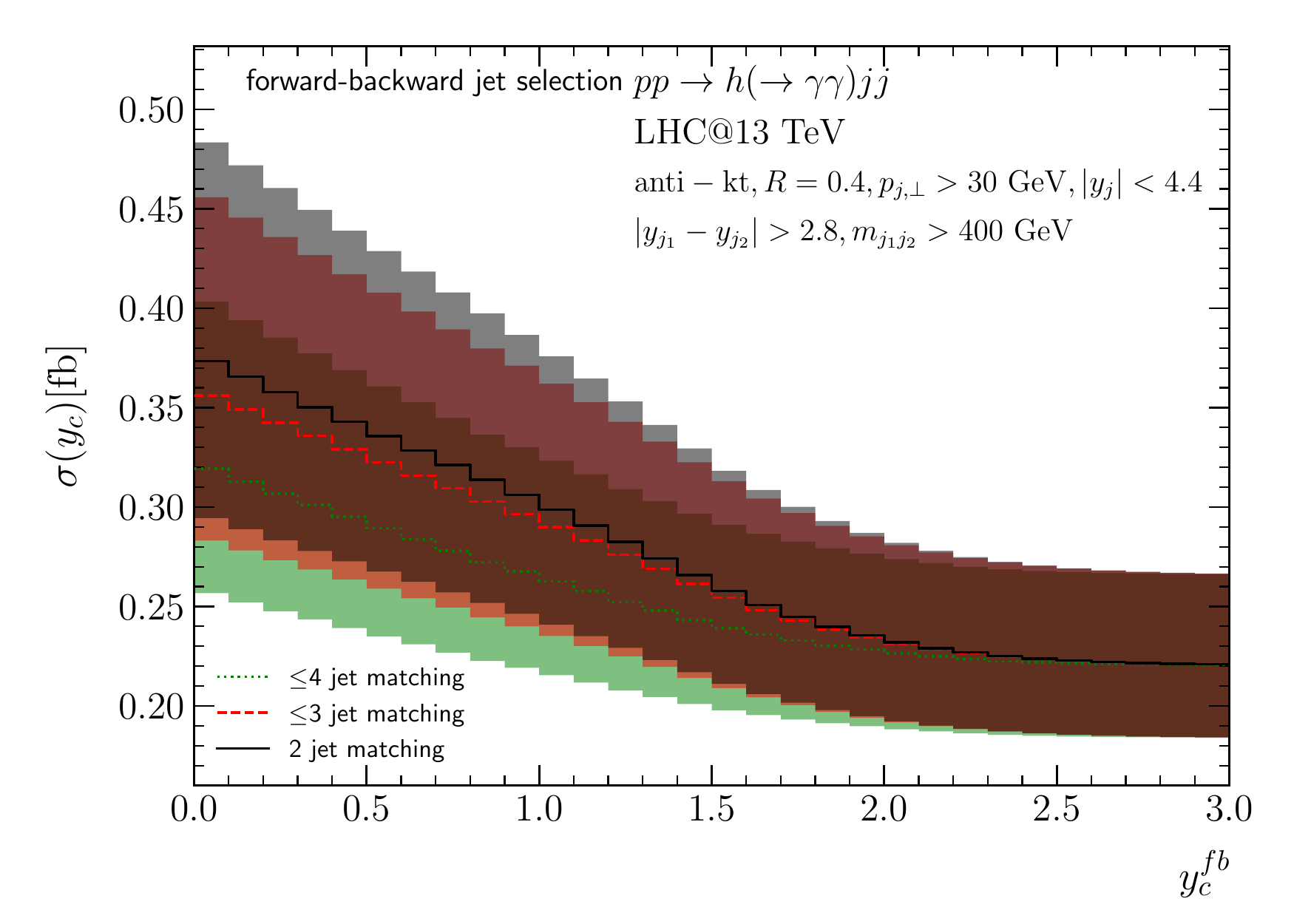}
  \caption{}
  \end{subfigure}
  \begin{subfigure}[t]{0.495\textwidth}
    \includegraphics[width=\linewidth]{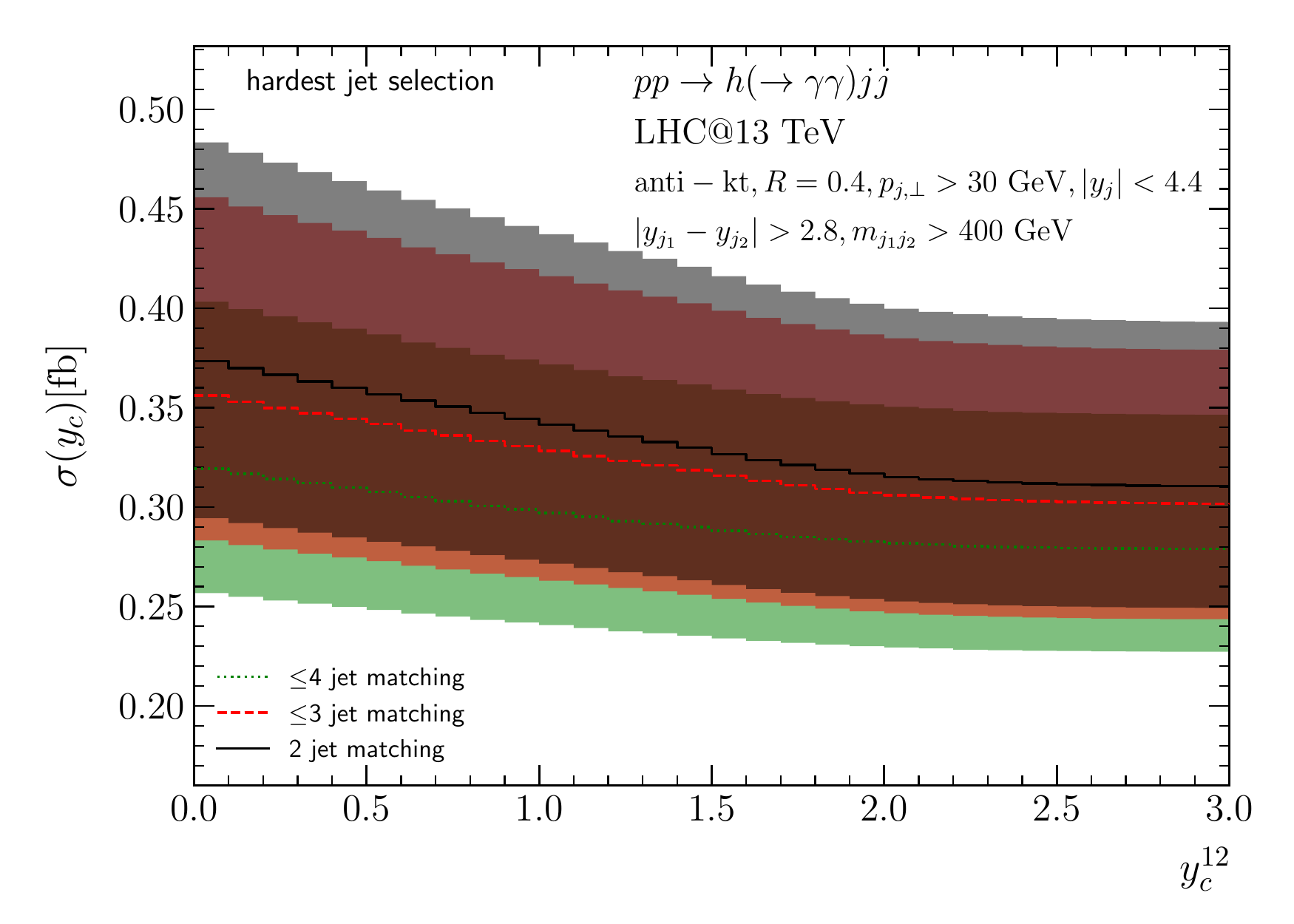}
    \caption{}
  \end{subfigure}
  \caption{Effect of a jet veto between (a) the most forward and
    backward jets and (b) the two hardest jets. Events with additional
    jets within a distance of $y_c$ to the rapidity centre are discarded.}
  \label{fig:yc}
\end{figure}

In order to achieve a greater reduction of the gluon-fusion background to weak-boson fusion
within the VBF-cuts, a veto on further jets can be applied. This has the added benefit of
reducing the contribution from higher jet multiplicities, which is
harder to predict in perturbation theory. The effectiveness
of such a cut relies on the difference in the
quantum corrections to the processes of VBF and
GF\cite{Dokshitzer:1991he}. Since this difference is due to the $t$-channel
colour-octet exchange of the GF process, we will apply a central jet veto
only in the regions away from the tagging jets, since the collinear regions
have similar emissions in VBF and GF. This is a slight improvement on the
normal central jet veto cuts, and is inspired by the Zeppenfeld variable\cite{Rainwater:1996ud}. Here, we consider a veto of
events with jets within a rapidity distance $y_c$ to the rapidity centre of either
(a) most forward and backward jets or (b) the hardest jets (see
also~\cite{Rainwater:1996ud,Bendavid:2018nar}). In case (b), we
only consider vetoing on further jets which are in between the two hardest jets. The results are
shown in Fig.~\ref{fig:yc}. As expected, the cross section in case (a)
converges for large $y_c$ to the exclusive
prediction for the production of a Higgs boson with exactly two jets,
irrespective of the fixed-order matching to higher multiplicities. When
applying the jet veto between the hardest jets in case (b), the overall
reduction in the GF component is much smaller, and insignificant compared to
the scale variation.

\subsection{Matching and Comparison to Fixed Next-to-Leading Order}
\label{sec:NLOcomp}
The complete reformulation of the formalism for matching and all-order
summation described in Sec.~\ref{sec:matching} has allowed for matching to
higher jet multiplicities in \HEJ. The impact of the four-jet matching on the
studied distributions is small. The method presented in the earlier sections
has been concerned with a point-by-point matching of the resummation to full
high-multiplicity tree-level accuracy. As extensively demonstrated in
Section~\ref{sec:res_4j}, this achieves perturbatively stable results for the
shapes of distributions. In order to reduce the scale variation and benefit
from the full NLO results for $Hjj$-production, we will now rescale the
results for \HEJ within the inclusive cuts of Eq.~\eqref{eq:photon_cuts} to
the NLO cross section for each choice of factorisation and renormalisation
scale. Thereby, full NLO accuracy is obtained for all dijet observables, LO
accuracy for trijet observables, and the impact on the shape of distributions
from four-jet contributions is accounted for at LO. This method was applied
also in Ref.\cite{Bendavid:2018nar}. While this approach does not change the
shape of distributions, the scale variation is reduced to the level of NLO
predictions.

We will here compare these predictions to those obtained at fixed NLO using
MCFM~\cite{Campbell:2006xx,Campbell:2010cz} and
SHERPA~\cite{Gleisberg:2008ta}.
\begin{figure}
  \centering
  \includegraphics[width=0.75\linewidth]{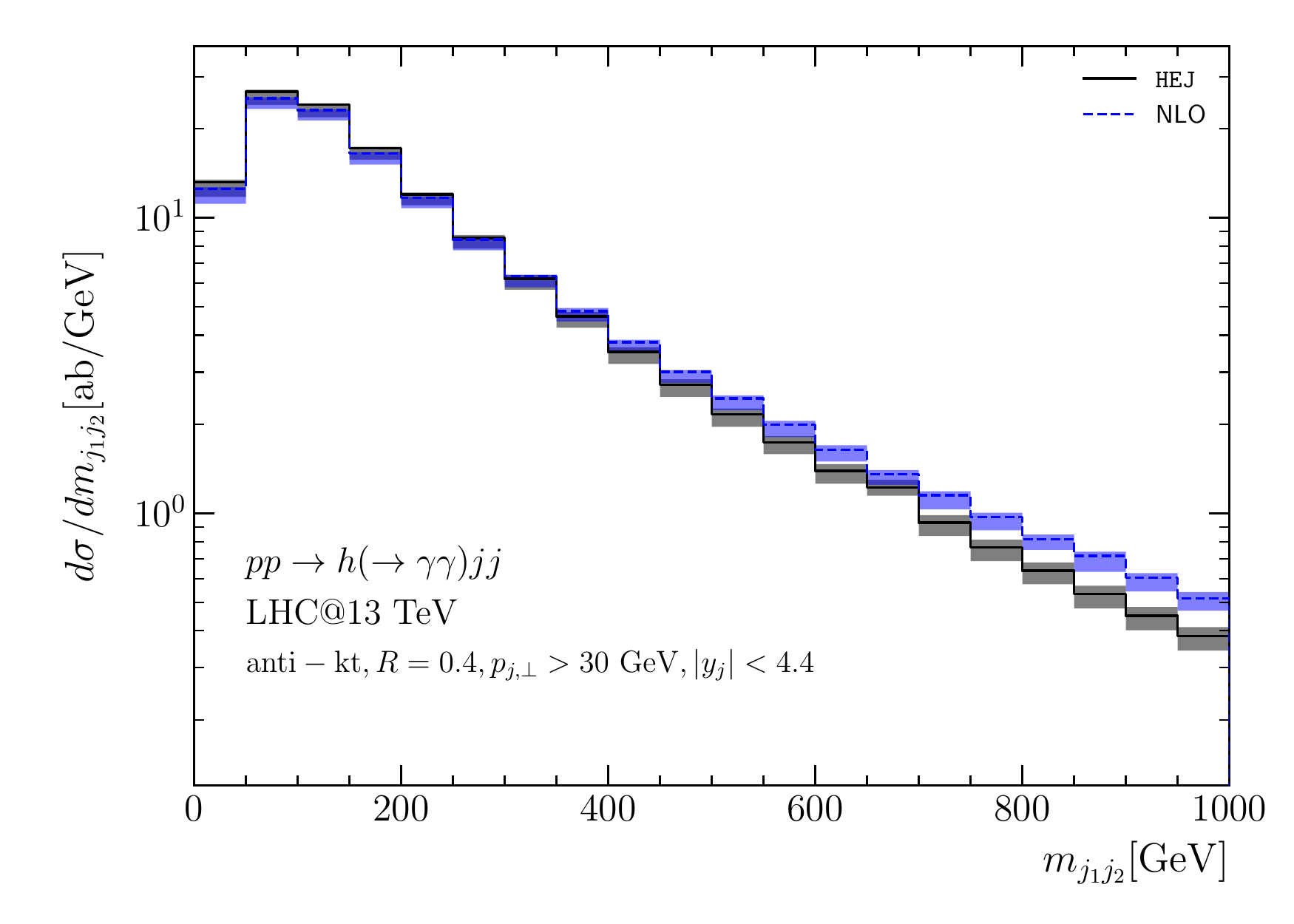}
  \caption{
    Distribution of the invariant mass between the two hardest
    jets. The \HEJ result with fixed-order matching up to 4
    jets is shown by the solid black line. The MCFM NLO prediction
    corresponds to the dashed blue line. The central renormalisation and factorisation
    scale is set to $H_T/2$, and they are varied independently by a factor of
    two.
  }
  \label{fig:NLO_cmp_minv}
\end{figure}
Fig.~\ref{fig:NLO_cmp_minv} compares the predictions for the distribution on
the invariant mass between the two hardest jets. The scale variation on the
\HEJ results is vastly reduced to that of Fig.~\ref{fig:m12}, as generally
expected by the inclusion of the full NLO corrections. The distribution
obtained with \HEJ for the invariant mass between the two hardest (in
transverse momentum) jets is still significantly steeper than that at pure
NLO, as a result of the possibility of significantly higher jet multiplicity,
and the fact that hard central jets have a slightly smaller PDF-suppression
than hard forward jets, and therefore the two hardest jets tend to also be
central. This means that the prediction for the cross section within the VBF
cuts is significantly smaller with \HEJ than for NLO, and indeed lies outside
the scale-variation band obtained at NLO.
\begin{figure}
  \centering
  \includegraphics[width=0.75\linewidth]{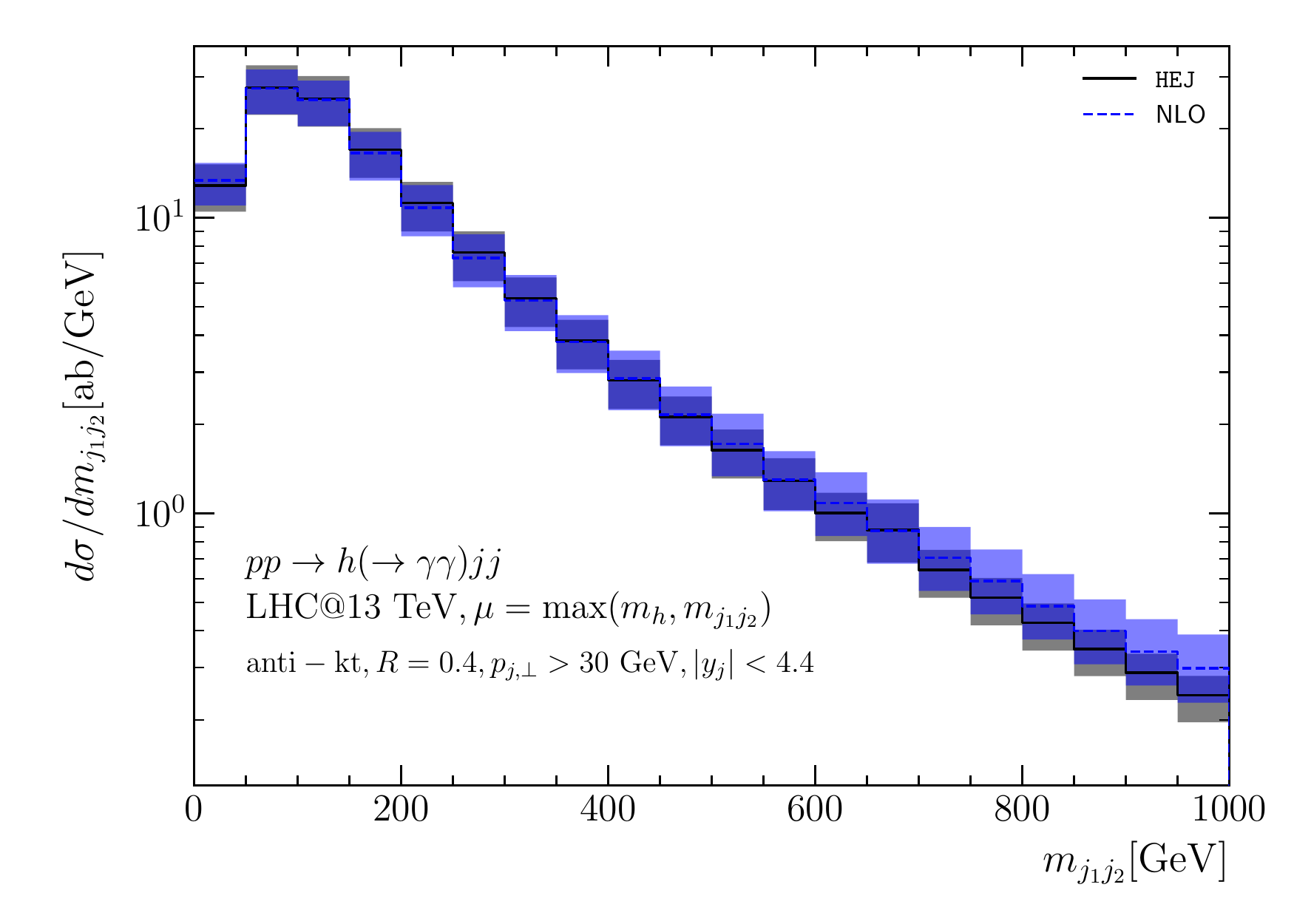}
  \caption{
    Distribution of the invariant mass between the two hardest
    jets, for a central scale choice of $m_{j_1 j_2}$. The \HEJ result with fixed-order matching up to 4
    jets is shown by the solid black line. The SHERPA NLO prediction
    corresponds to the dashed blue line.
  }
  \label{fig:NLO_cmp_minv_m12}
\end{figure}
In numbers, the cross sections obtained (at NLO) for
$pp\to h(\to \gamma\gamma)jj$ for inclusive cuts and with a central scale
choice of $\mu_r=\mu_f=H_T/2$ is $6.58^{+0.08}_{-0.57}$~fb. This is obviously
the same as that obtained with \HEJ for inclusive cuts, once the cross
sections are normalised to NLO accuracy. For the VBF cuts, the NLO cross
section is $0.872^{+0.024}_{-0.090}$~fb, and that obtained for \HEJ is
$0.561^{+0.031}_{-0.067}$~fb. Even though the inclusive cross section for
\HEJ is normalised to that obtained at NLO, a sizeable difference in the
cross section within the VBF cuts arises due to a difference in the slope of
distribution in $m_{j_1 j_2}$ and the requirement of $m_{j_1 j_2}>400$~GeV
for the VBF cuts. The VBF cuts cause a similar reduction in the cross section to
$13.2\%$ (NLO) and $8.5\%$ (\HEJ) of the inclusive cross section respectively.

Comparing the results of
Fig.~\ref{fig:NLO_cmp_minv} and Fig.~\ref{fig:NLO_cmp_minv_m12} we observe
that a choice of a central scale for the NLO calculation of
$\mu_r=\mu_f=H_T/2$ leads to a suspiciously small scale variation - and
indeed the central scale choice gives results close to the extremum
obtained with the variations, despite the scales being varied either side of
the central choice of $\mu_r=\mu_f=H_T/2$. Such a behaviour of the scale
variation often indicates that the NLO scale variation obtained with this
scale choice is underestimating the theoretical
uncertainty\cite{Currie:2017eqf}.

Indeed, Ref.\cite{Currie:2017eqf} investigated the distribution in
$m_{j_1 j_2}$ for dijet production at NNLO at the LHC, and found that at
large $m_{j_1 j_2}$ this scale choice is favoured over $p_T$ based on
perturbative convergence. The invariant mass between the two hardest jets
obviously is not a stable perturbative scale choice for all bins in the
distribution, which extends to very low values of $m_{j_1 j_2}$.  With a
central scale choice of $\mu_f=\mu_r=\max(m_h,m_{j_1 j_2})$, the central
scale choice leads to predictions in the centre of the variation band. The
scale variation bands obtained with NLO and \HEJ also overlaps in each
bin of the distribution. With this central scale choice, the cross sections
obtained at NLO for inclusive cuts is $6.23^{+1.11}_{-1.22}$~fb. For the VBF
cuts, the NLO cross section is $0.542^{+0.156}_{-0.125}$~fb, and that obtained
for \HEJ is $0.359^{+0.045}_{-0.061}$~fb. The VBF cuts cause a similar
reduction in the cross section to 8.7\% (NLO) and 5.8\% (\HEJ) of the
inclusive cross section respectively.

It is worth noting that (ignoring the mass of each jet) since
$m^2_{j_1 j_2}=2 p_{\perp j_1} p_{\perp j_2} (\cosh(y_{j_1}-y_{j_2}) -
\cos(\phi_{j_1} -\phi_{j_2}))$ a central scale choice of $\mu_r=m_{j_1 j_2}$
systematically runs $\alpha_s$ such that $\alpha_s \Delta y_{j_1 j_2}$ tends
to a constant for large $\Delta y_{j_1 j_2}$. This would seem to spoil the
standard argument of BFKL noting large and systematic leading logarithmic
corrections of the form $(\alpha_s \Delta y_{j_f j_b})^k$ at large
$\Delta y_{j_f j_b}$, at least for $\Delta y_{j_1 j_2}$ sufficiently large
that $m_{j_1 j_2}$ is close to the hadronic collision energy that only two
jets exists, because hard radiation beyond the two jets required is
suppressed. For events with more than two jets, there is no direct
correlation between $\Delta y_{j_1 j_2}$ and $\Delta y_{j_f j_b}$. The
results for the scale choice of $\mu_r=\mu_f=m_{j_1 j_2}$ are discussed
further in appendix~\ref{sec:invmass}. Here we just note that the apparent
convergence of the perturbative series (i.e.~a comparison of the LO and NLO
results and scale variation) is not significantly different for the two scale
choices. The scale variation around $\mu_r=\mu_f=H_T/2$ is accidentally
small, since the central scale choice leads to the maximum cross section
within the variation.

\begin{figure}
  \centering
  \begin{subfigure}[t]{0.495\textwidth}
    \includegraphics[width=\linewidth]{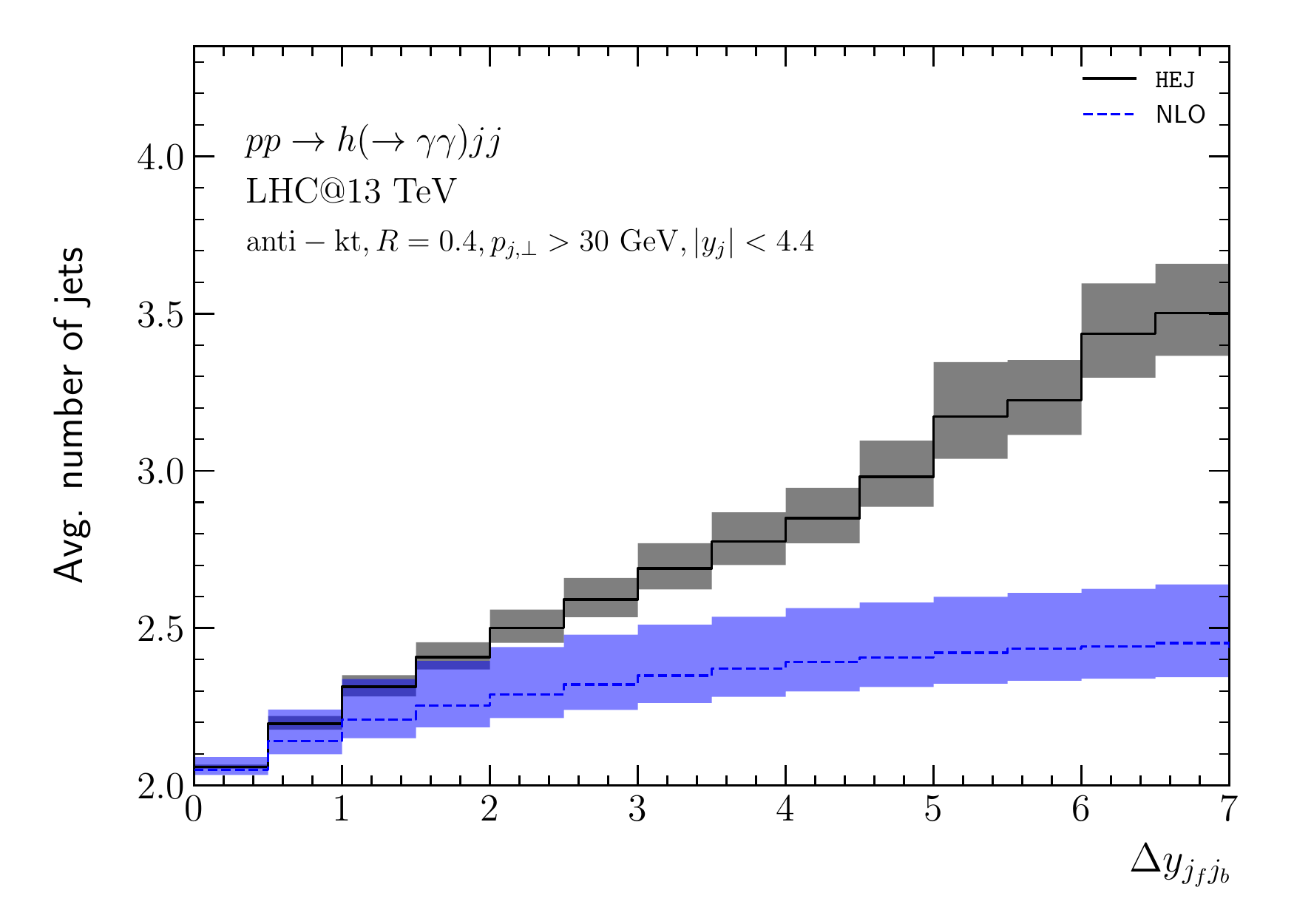}
    \caption{}
  \end{subfigure}
  \begin{subfigure}[t]{0.495\textwidth}
    \includegraphics[width=\linewidth]{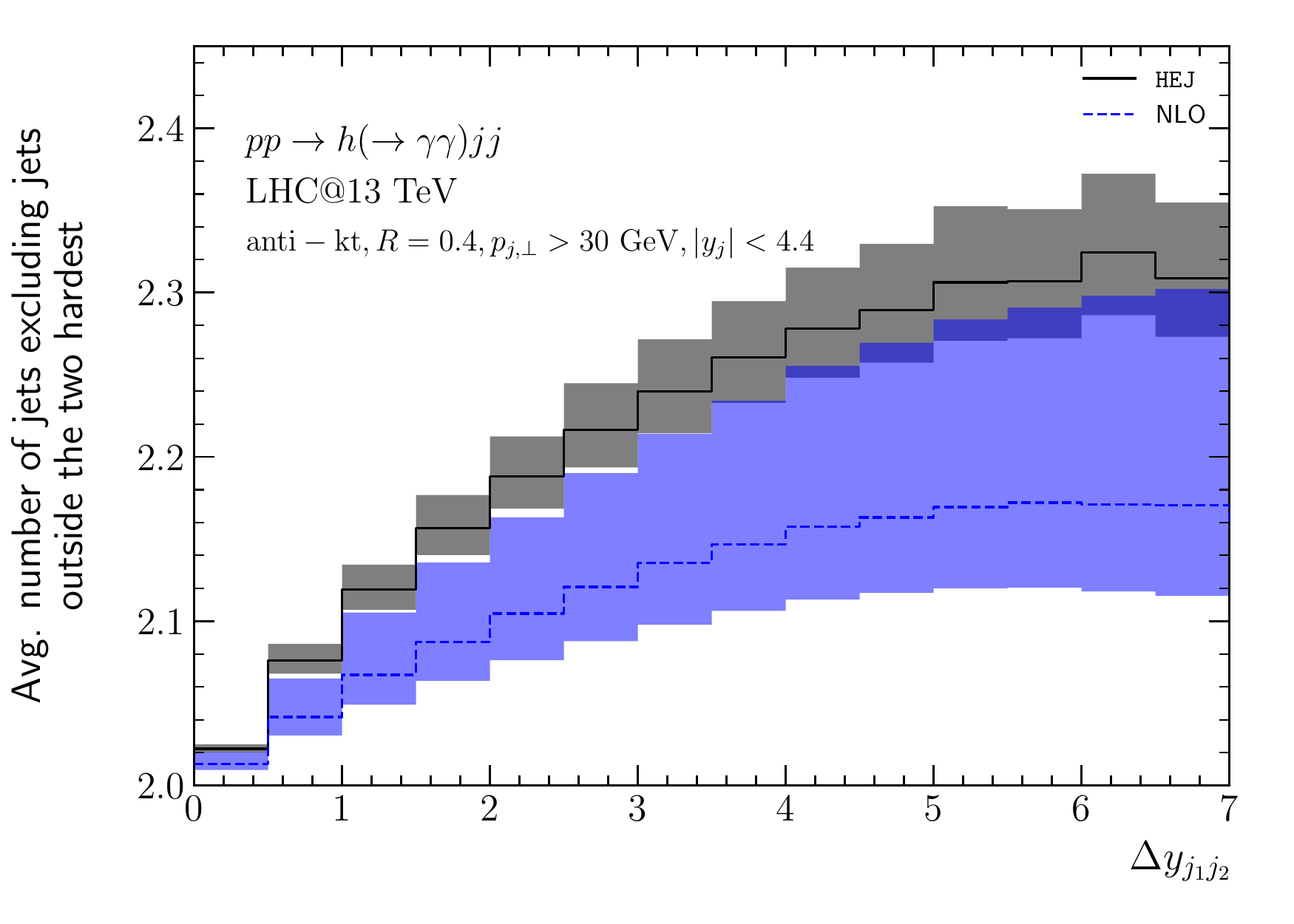}
    \caption{}
  \end{subfigure}
  \begin{subfigure}[t]{0.495\textwidth}
    \includegraphics[width=\linewidth]{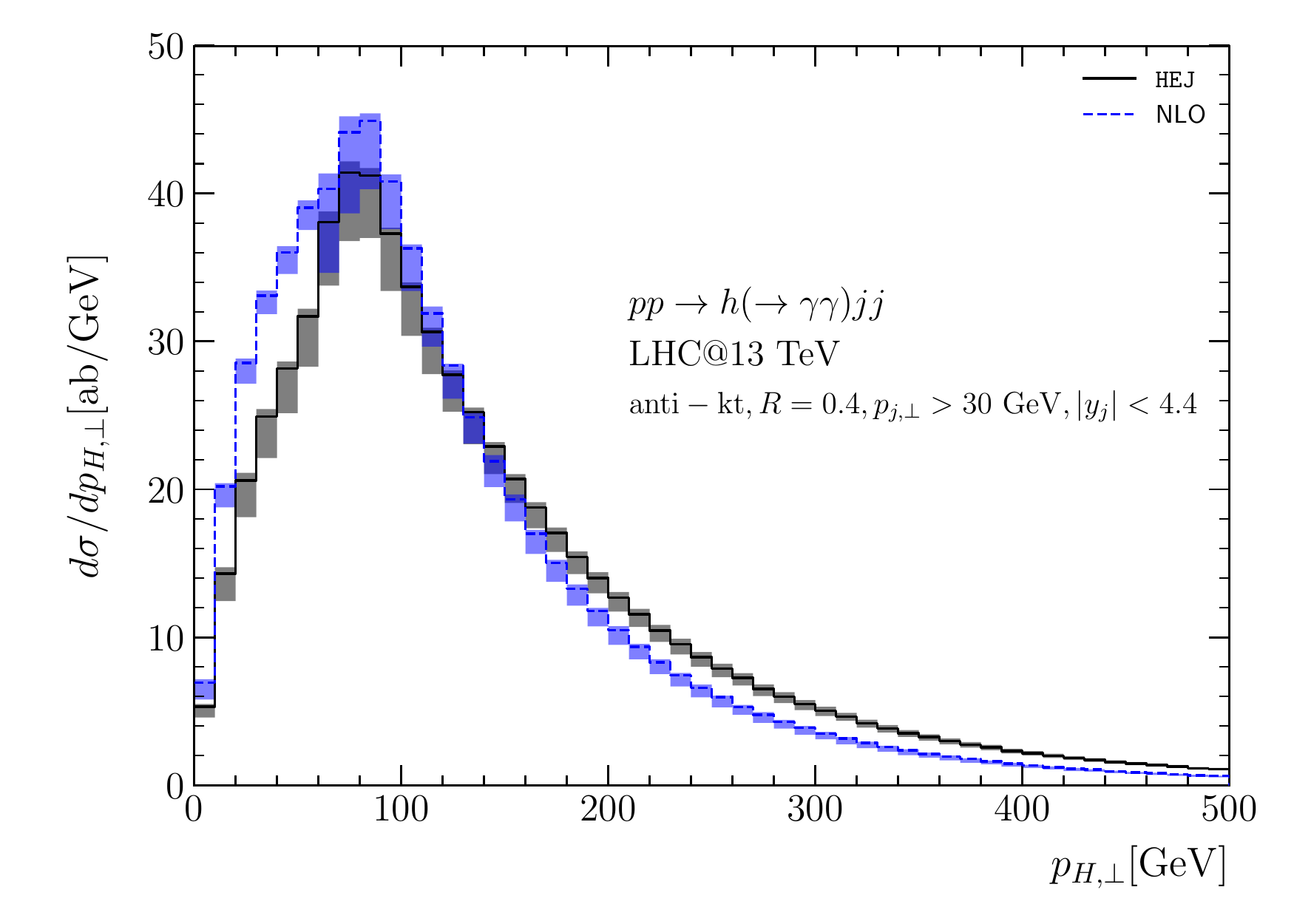}
    \caption{}
  \end{subfigure}
  \begin{subfigure}[t]{0.495\textwidth}
    \includegraphics[width=\linewidth]{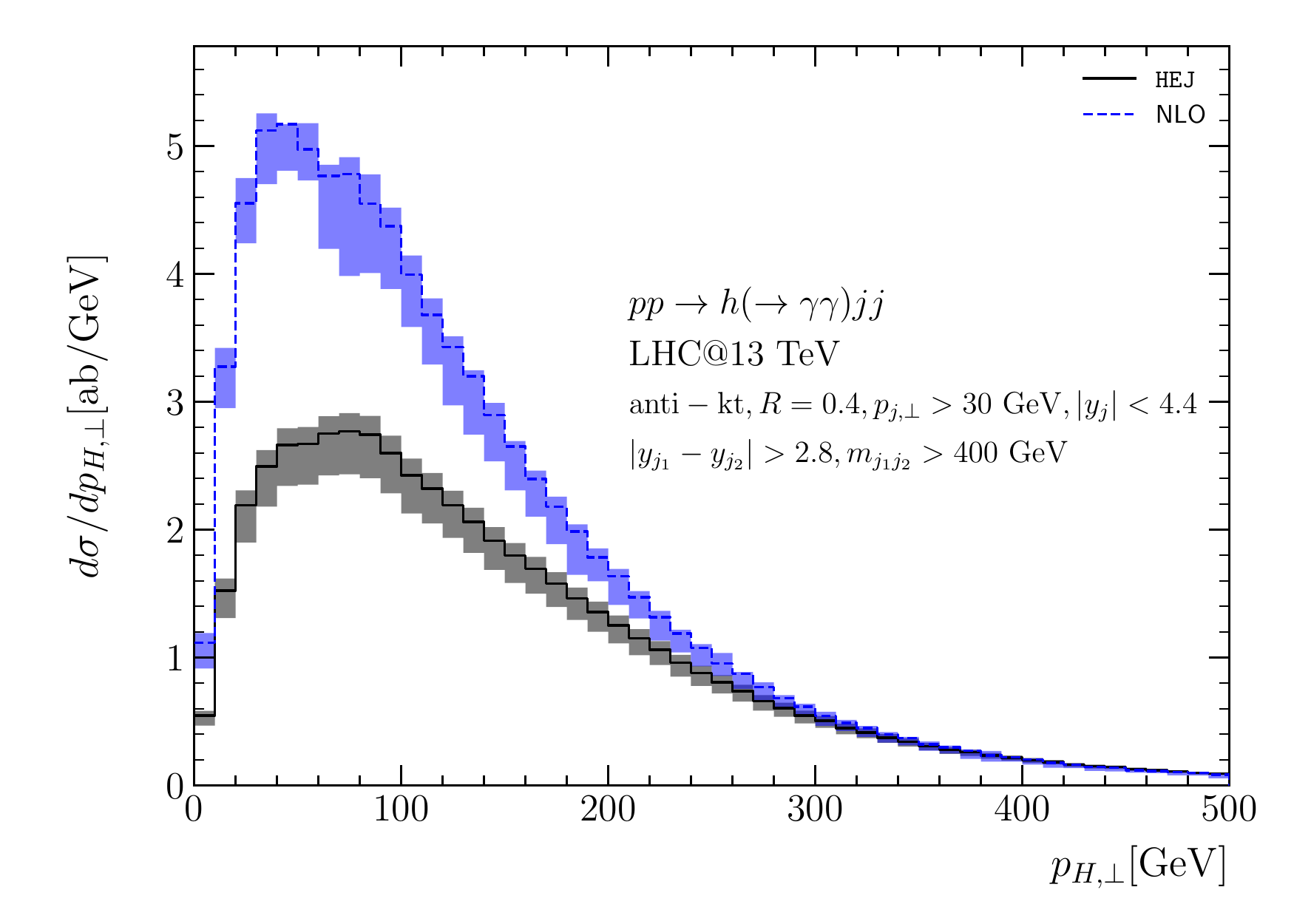}
    \caption{}
  \end{subfigure}
  \begin{subfigure}[t]{0.495\textwidth}
    \includegraphics[width=\linewidth]{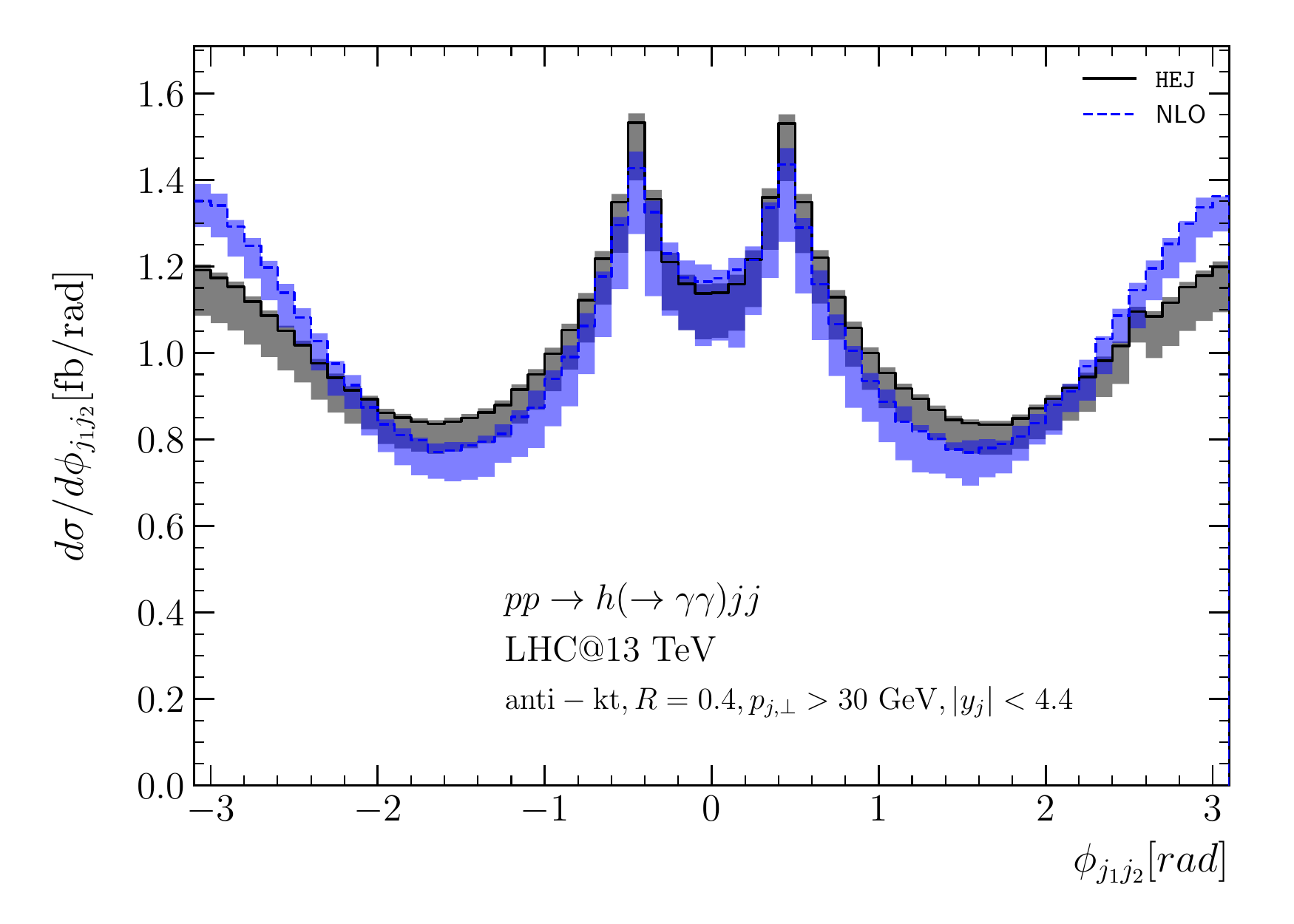}
    \caption{}
  \end{subfigure}
  \begin{subfigure}[t]{0.495\textwidth}
    \includegraphics[width=\linewidth]{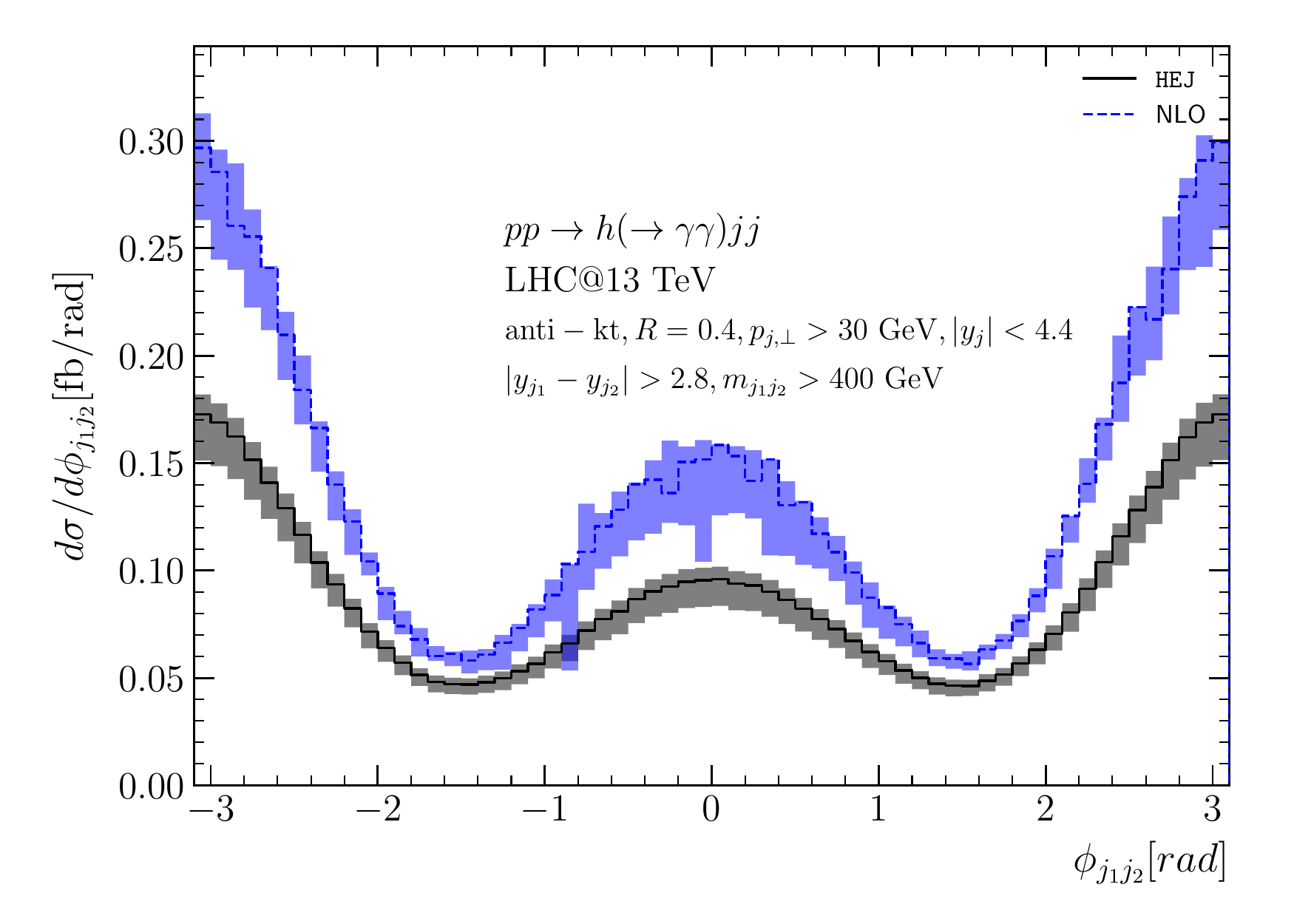}
    \caption{}
  \end{subfigure}
  \caption{Comparison of \HEJ results with fixed-order matching up to 4
jets (solid black line) with NLO predictions from MCFM (dashed blue
line). The shown observables are (a) the average jet multiplicity, (b)
the number of jets in between the two hardest jets, the
distribution of the Higgs-boson transverse momentum with (c) inclusive and
(d) VBF cuts, and the distribution of the azimuthal angle between
the hardest jets with (e) inclusive and (f) VBF cuts.}
  \label{fig:NLO_cmp}
\end{figure}
Fig.~\ref{fig:NLO_cmp}(a) and Fig.~\ref{fig:NLO_cmp}(b) investigates the
potential for using perturbative corrections in the form of additional
jet-radiation as a means of identifying the gluon-fusion production
channel. For the same event selection, the figure compares the results for
the average number of jets counting additional jets (a) between the most
forward and backward jets, and (b) in-between the two hardest jets only.  The
results on Fig.~\ref{fig:NLO_cmp}(a) are relevant for e.g.~jet vetos between
the most forward and most backward hard jet, whereas Fig.~\ref{fig:NLO_cmp}(b)
is relevant if the veto is applied between just the two hardest jets in the
event. The results for \HEJ are identical to those for 4-jet matching in
Fig.~\ref{fig:njets} (since just the total cross section has been adjusted to
the NLO result for $Hjj$), but the results are here compared to those
obtained using the NLO calculation for $Hjj$-production. As observed also in
previous analyses\cite{Campbell:2013qaa}, the results obtained at NLO tends
towards 2.5, where the exclusive, hard three-jet cross section is as large as
the two-jet cross section. This clearly illustrates the slow convergence of
the perturbative series. The results for NLO and for \HEJ start diverging
already at small $\Delta y_{j_f j_b}$. It is worth noting that the linear
growth in the number of hard jets vs.~$\Delta y_{j_f j_b}$ has been
experimentally confirmed\cite{Aad:2011jz,Abazov:2013gpa} for several
processes with colour octet exchanges in the $t$-channel.

Even though exactly the same events are involved, the breakdown of the
convergence is less obvious in Fig.~\ref{fig:NLO_cmp}(b). The number of jets
in-between the two hardest is obviously smaller, and both the results for
\HEJ and for NLO appear to asymptote to a value for the average number of
jets of 2.2 for NLO and 2.3 for \HEJ.

Fig.~\ref{fig:NLO_cmp} also shows the predictions for the Higgs transverse
momentum spectrum obtained at NLO and with \HEJ both for inclusive (c) and
VBF-cuts (d). The distributions are very similar for inclusive cuts, with a
peak around 80~GeV, and the spectrum from \HEJ slightly harder. For VBF cuts,
the prediction for \HEJ is lower than that for NLO, as a result of the
steeper spectrum in $m_{j_1 j_2}$ and the requirement of
$m_{j_1 j_2}>400$~GeV. The two predictions for the high-$p_\perp$ tail within
the VBF cuts coincide, but in this region the infinite top-mass approximation
is certainly not trustworthy.

Finally, Fig.~\ref{fig:NLO_cmp}(c) and (d) compares the azimuthal angle
between the two hardest jets for (c) inclusive and (d) VBF cuts
respectively. In both the distributions, the region of back-to-back jets at
$\phi=\pm\pi$ is slightly suppressed in \HEJ compared to NLO. The valley at
$\phi=0$ for the inclusive cut is due to the jet-algorithm removing the collinear
region. The result within the VBF cuts on Fig.~\ref{fig:NLO_cmp}(f) show that
the reduction in the cross section within the VBF cuts for \HEJ compared to
NLO predominantly is in the region of back-to-back jets and jets in the same
azimuthal direction. The region at $\phi_{j_1 j_2}=0$ is not collinear within
the VBF cuts, and so the structure induced by the jet algorithm within the
inclusive cuts of Fig.~\ref{fig:NLO_cmp}(e) is not present within the VBF
cuts of Fig.~\ref{fig:NLO_cmp}(f).


\section{Conclusions}
\label{sec:summary}
We have presented a reformulation of the matching formalism within \HEJ,
which recasts the calculation as one of merging fixed-order samples of
increasing multiplicity. The merging is performed respecting the resummation
of perturbative terms logarithmically enhanced at large $\hat s/p_t^2$. While
the formalism is mathematically equivalent to that previously used, stable
results are obtained using orders of magnitudes less CPU time. This allows
matching to be performed to higher multiplicity.

The new formalism was used in a study of Higgs-boson production in
association with dijets. The impact of the higher-multiplicity merging is
minimal on the shape of distributions important for the application of VBF
cuts. For a central scale choice of $\mu_f=\mu_r=H_T/2$, the VBF cuts reduce
the inclusive cross section of $6.58^{+0.08}_{-0.57}$~fb on the
$h\to\gamma\gamma$-channel to 13\% ($0.872^{+0.024}_{-0.090}$~fb) at NLO, or
8.5\% ($0.561^{+0.031}_{-0.067}$~fb) once both NLO and the \HEJ-corrections
are accounted for. The further suppression within \HEJ is due to a steeper
falling spectrum in the invariant mass between the two hardest jets. The NLO
scale dependence is estimated by varying the renormalisation and
factorisation scale independently by a factor of two. However, the scale
variation around $H_T/2$ is artificially small, since the central scale
choice achieves a value close to the maximum within the variations. With a
scale choice of $\mu_r=\mu_f=m_{j_1 j_2}$ (but bounded from below by $m_h$),
the spectrum is similar at NLO and with the further \HEJ-corrections. With
the scale choice, the inclusive NLO cross section for
$h(\to \gamma\gamma)jj$-cross section is $6.23^{+1.11}_{-1.22}$~fb, and
$0.542^{+0.156}_{-0.125}$~fb within the VBF cuts. The result for \HEJ within
the VBF cuts is $0.359^{+0.045}_{-0.061}$~fb. The VBF cuts cause a similar
reduction in the cross section to 8.7\% (NLO) and 5.8\% (\HEJ) of the
inclusive cross section respectively.

The formalism presented in this report will be instrumental in the further
developments, including an account of heavy quark mass effects, and mergings
of NLO cross sections.


\section*{Acknowledgements}
The authors would like to thank Gavin Salam for discussions on the anti-kt
jet clustering algorithm.

This work has received funding from the European Union's Horizon 2020
research and innovation programme as part of the Marie Sk\l{}odowska-Curie
Innovative Training Network MCnetITN3 (grant agreement no. 722104) and UK
Science and Technology Facilities Council (STFC). JMS is supported by a Royal
Society University Research Fellowship and the ERC Starting Grant 715049
``QCDforfuture''.


\appendix
\boldmath
\section{Result for the Central Scale Choice of $m_{j_1 j_2}$}
\begin{figure}
  \centering
  \begin{subfigure}[t]{0.495\textwidth}
    \includegraphics[width=\linewidth]{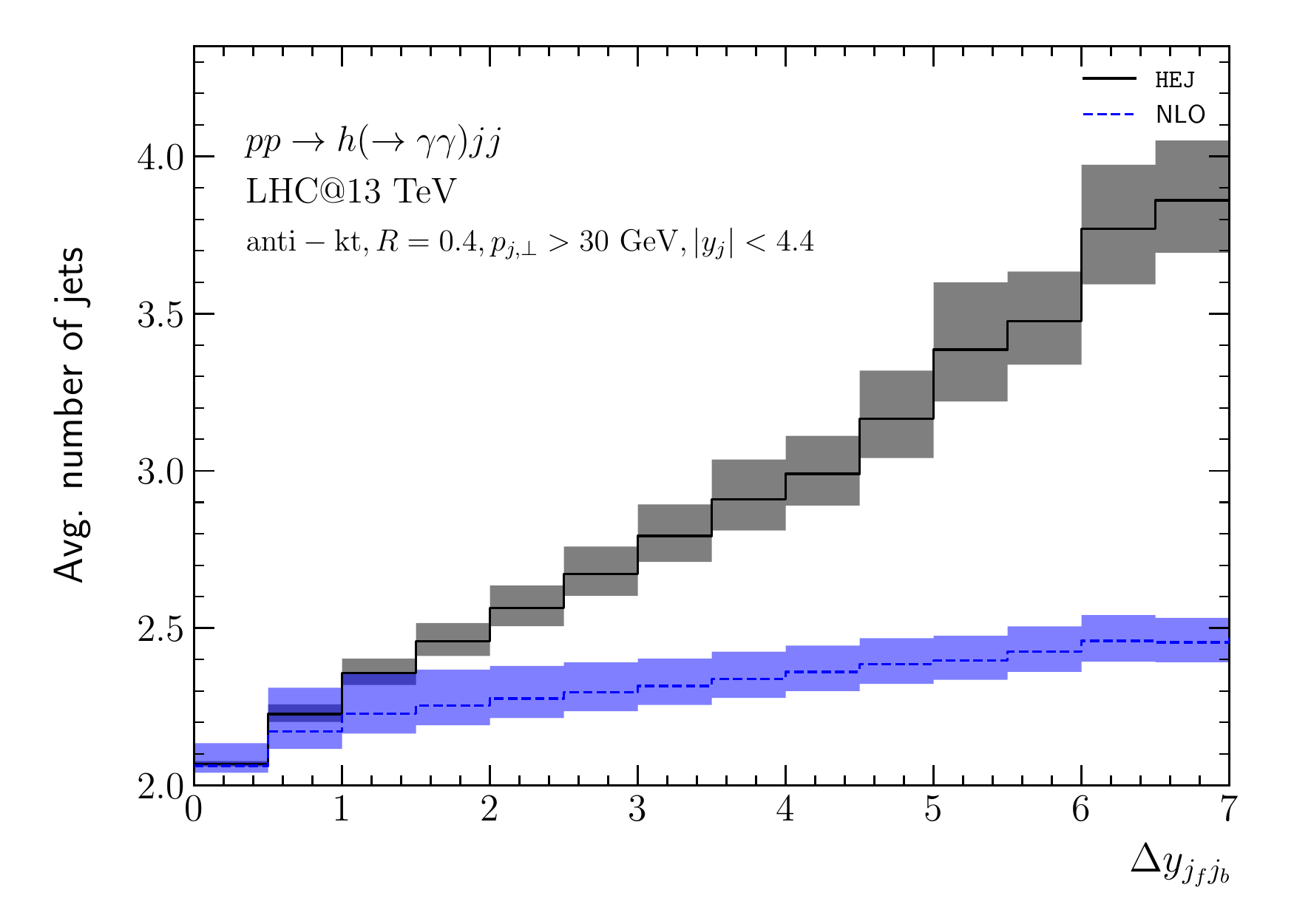}
    \caption{}
  \end{subfigure}
  \begin{subfigure}[t]{0.495\textwidth}
    \includegraphics[width=\linewidth]{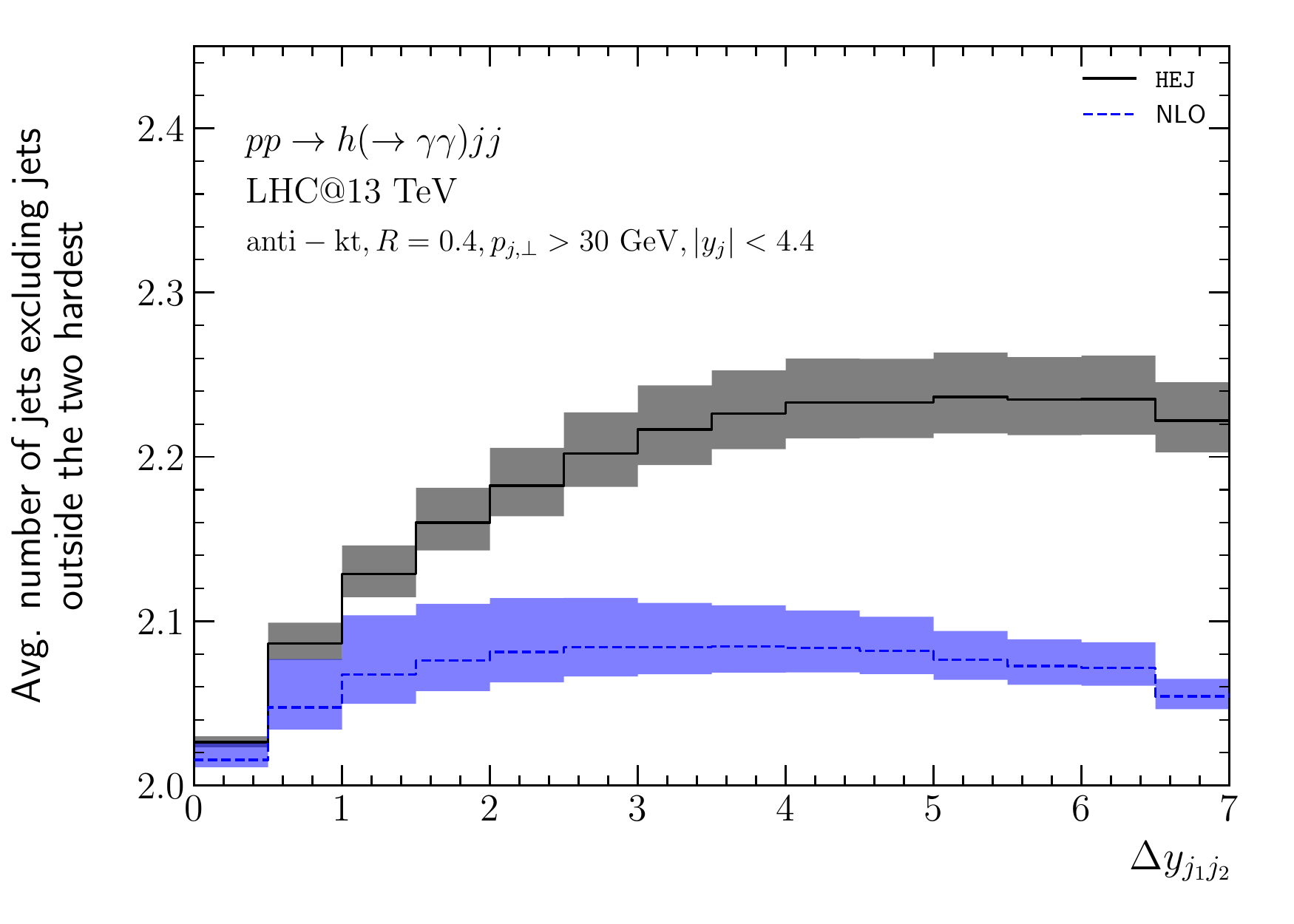}
    \caption{}
  \end{subfigure}
  \begin{subfigure}[t]{0.495\textwidth}
    \includegraphics[width=\linewidth]{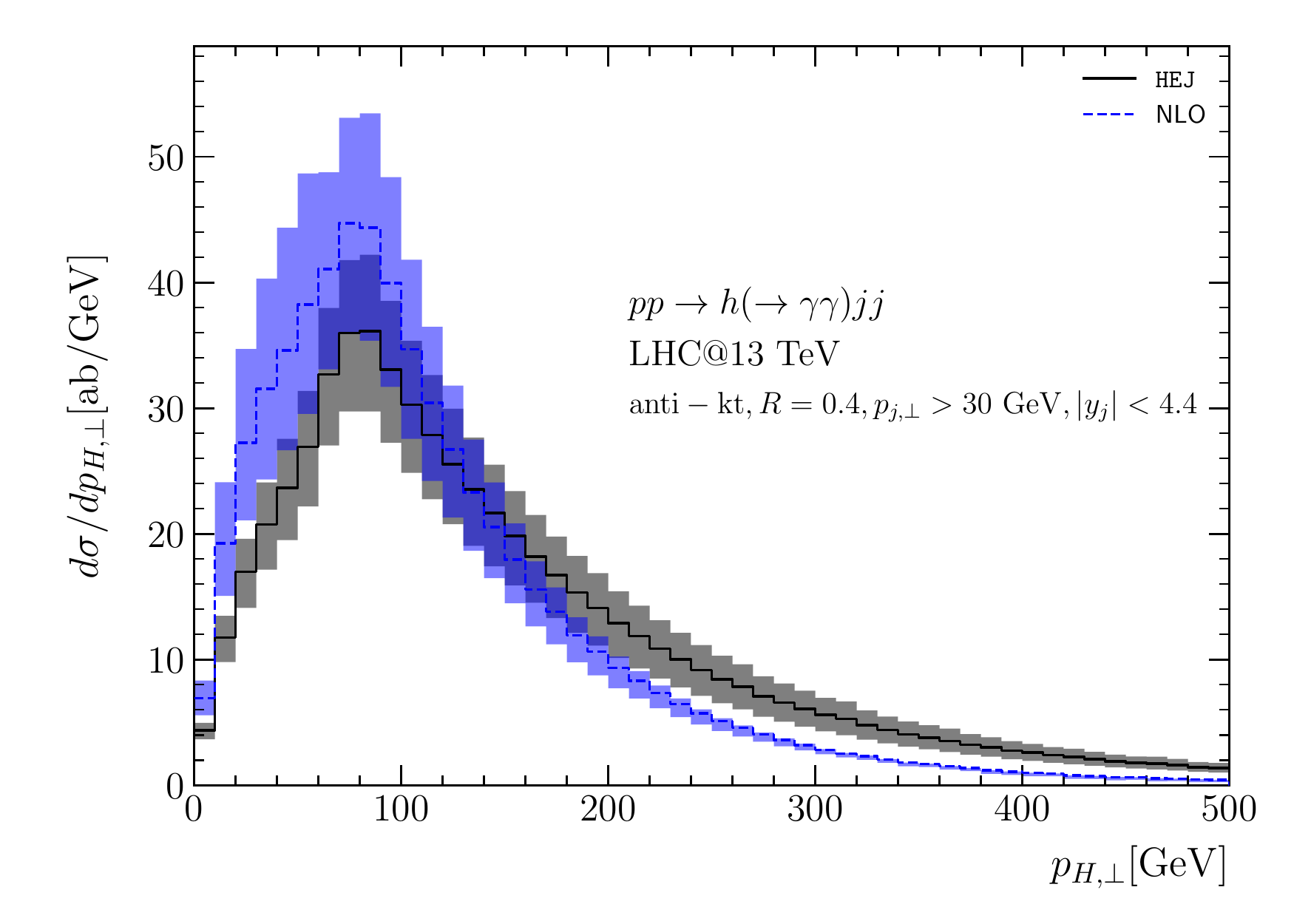}
    \caption{}
  \end{subfigure}
  \begin{subfigure}[t]{0.495\textwidth}
    \includegraphics[width=\linewidth]{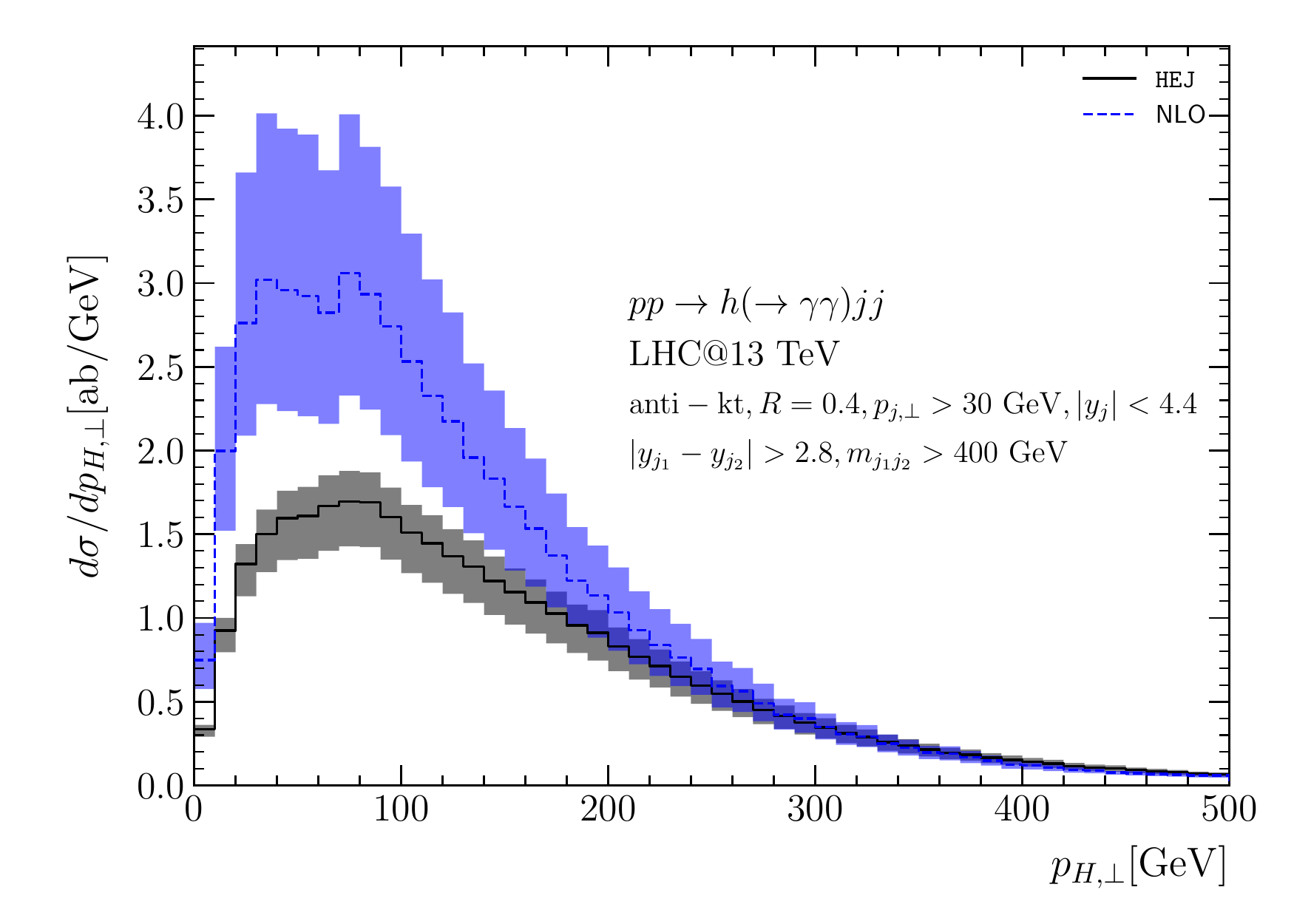}
    \caption{}
  \end{subfigure}
  \begin{subfigure}[t]{0.495\textwidth}
    \includegraphics[width=\linewidth]{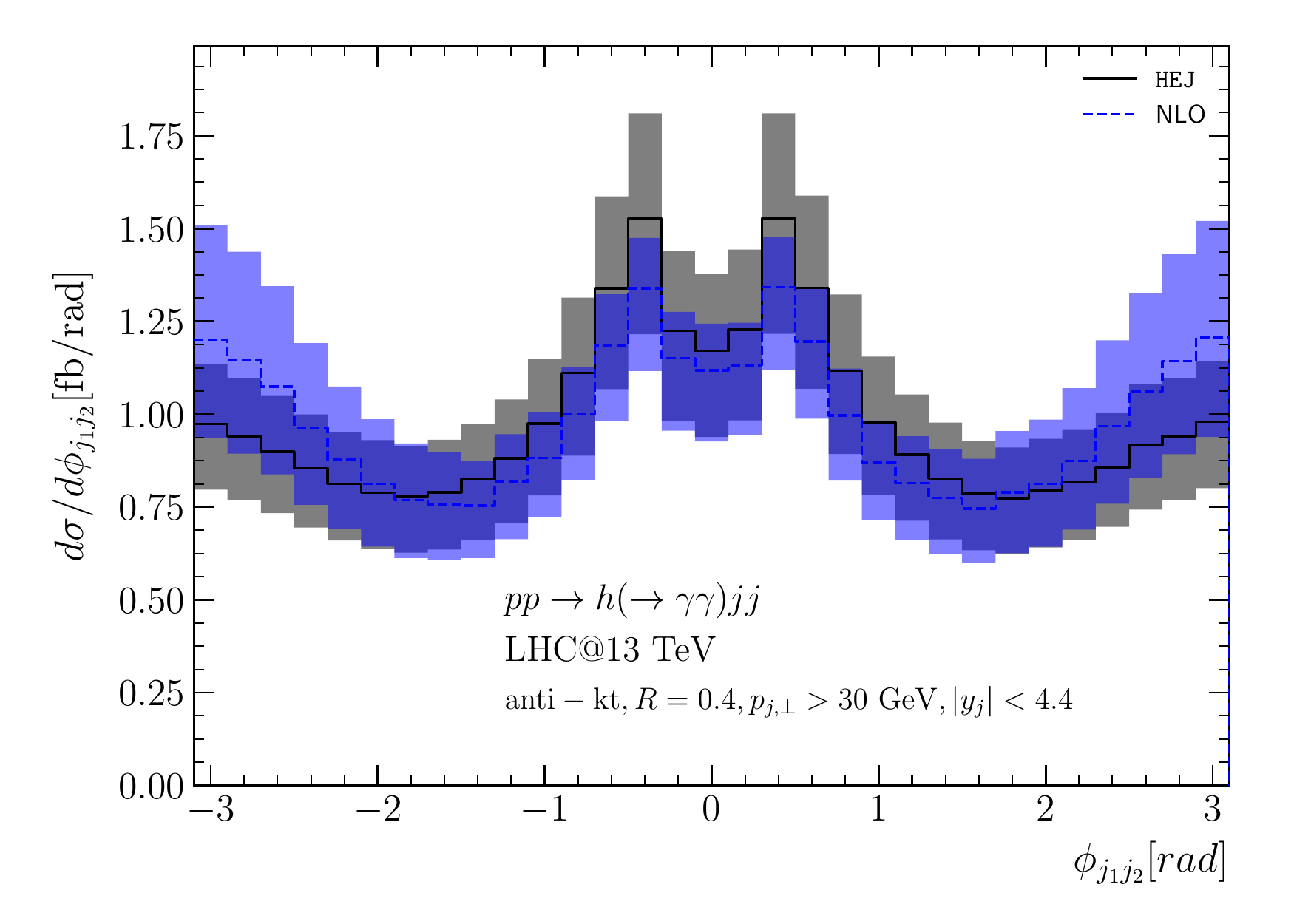}
    \caption{}
  \end{subfigure}
  \begin{subfigure}[t]{0.495\textwidth}
    \includegraphics[width=\linewidth]{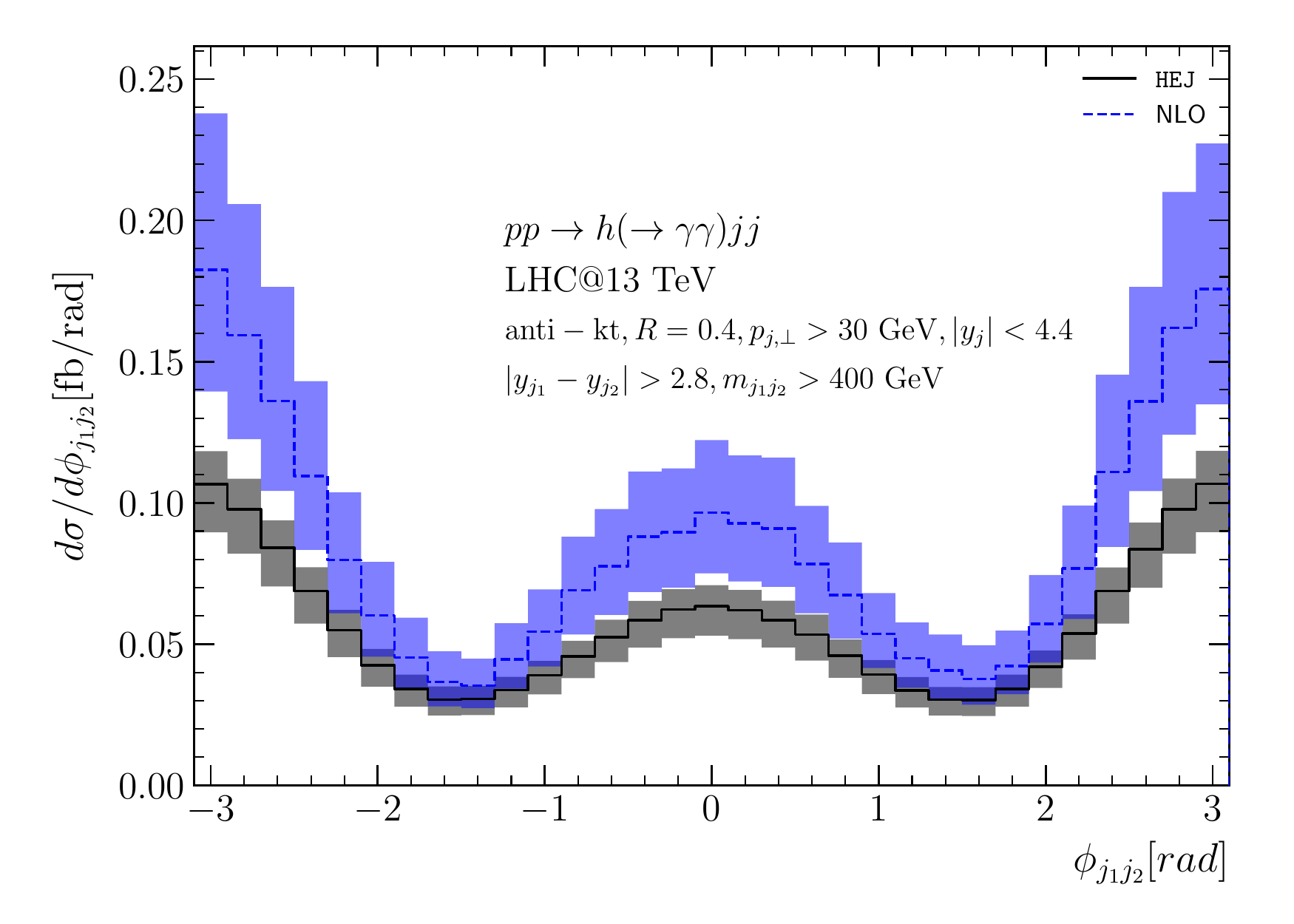}
    \caption{}
  \end{subfigure}
  \caption{Comparison of \HEJ results with fixed-order matching up to 4
jets (solid black line) with NLO predictions from SHERPA (dashed blue
line) for a central scale choice of $\mu_r=\mu_f=\max(m_h,m_{j_1 j_2})$. The shown observables are (a) the average jet multiplicity, (b)
the number of jets in between the two hardest jets, the
distribution of the Higgs-boson transverse momentum with (c) inclusive and
(d) VBF cuts, and the distribution of the azimuthal angle between
the hardest jets with (e) inclusive and (f) VBF cuts.}
  \label{fig:NLO_cmp_m12}
\end{figure}

\unboldmath
\label{sec:invmass}
Fig.~\ref{fig:NLO_cmp_m12} shows the same distributions for NLO and \HEJ
(with the inclusive cross section scaled to that of NLO for each scale choice)
as those investigated on Fig.~\ref{fig:NLO_cmp}, but for predictions obtained
using a central scale of $\mu_f=\mu_r=\max(m_h,m_{j_1 j_2})$. Panel (a) shows the
average number of hard jets vs.~the rapidity difference between the
forward-backward jet pair. The result for NLO is similar to that obtained with the
scale $H_T/2$, but the NLO scale variation is \emph{reduced} (even though the
scale variation on the NLO cross section themselves is increased by using
$m_{j_1 j_2}$ instead of $H_T/2$). The rise in the number of hard jets is
slightly stronger for \HEJ than with the scale of $H_T/2$.

Fig.~\ref{fig:NLO_cmp_m12}(b) shows the average number of jets, counting only
additional jets if their rapidity is in-between that of the two hardest
jets. The behaviour of the NLO prediction for small $\Delta y_{j_1 j_2}$ is
similar to that displayed on Fig.~\ref{fig:NLO_cmp}(b) until
$\Delta y_{j_1 j_2}\sim 3$, after which the average number of jets
\emph{decreases}. This behaviour is a result of the decreasing value of
$\alpha_s$ for large $\Delta y_{j_1 j_2}$ with this scale choice.

The remaining plots on Fig.~\ref{fig:NLO_cmp_m12}(c)-(f) all show similar
features to those of Fig.~\ref{fig:NLO_cmp}, but with a smaller cross section
and larger scale variation for the results at NLO.

\newpage
\bibliographystyle{JHEP}
\bibliography{papers}

\end{document}